\documentclass[english, 12pt]{article}

\ifx\pdftexversion\undefined\else
	\pdfoutput=1
	\usepackage[T1]{fontenc}
	\usepackage[utf8]{inputenc}
\fi

\usepackage{microtype}
\usepackage{babel}
\usepackage{lmodern}
\usepackage{csquotes}
\usepackage{fullpage}
\usepackage{zref-clever}
	\newcommand\cref{\zcref}
	\zcsetup{abbrev=true}
\usepackage{mathtools}
	\mathtoolsset{showonlyrefs}
	\numberwithin{equation}{section}
\usepackage{amsthm}
\usepackage{amssymb}
\usepackage{mathrsfs}
\usepackage[hidelinks]{hyperref}
\usepackage[citestyle=numeric-comp, giveninits, backref]{biblatex}
\usepackage{tikz}
\usepackage{tabu}
\usepackage{authblk}
	
\newtheorem{lemma}{Lemma}[section]
\newtheorem{theorem}[lemma]{Theorem}
\newtheorem{conjecture}[lemma]{Conjecture}
\theoremstyle{remark}
\newtheorem{remark}[lemma]{Remark}
\theoremstyle{definition}
\newtheorem{definition}[lemma]{Definition}

\AddToHook{env/theorem/begin}{%
    \zcsetup{countertype={lemma=theorem}}}
\AddToHook{env/conjecture/begin}{%
    \zcsetup{countertype={lemma=conjecture}}}
\AddToHook{env/remark/begin}{%
    \zcsetup{countertype={lemma=remark}}}
\AddToHook{env/definition/begin}{%
    \zcsetup{countertype={lemma=definition}}}
\zcRefTypeSetup{conjecture}{
Name-sg = Conjecture ,
name-sg = conjecture ,
Name-pl = Conjectures ,
name-pl = conjectures ,
}

\DeclareMathOperator\tr{tr} 
\DeclareMathOperator\rk{rk} 
\DeclareMathOperator\im{im}

\DeclareMathOperator\End{End} 
\DeclareMathOperator\Hom{Hom} 
\DeclareMathOperator\ch{ch}
\DeclareMathOperator\td{td}
\DeclareMathOperator\Lie{Lie}

\newcommand\ii{\mathrm i} 
\newcommand\BZ{\mathbb Z} 
\newcommand\BC{\mathbb C} 
\newcommand\BR{\mathbb R} 
\newcommand\BP{\mathbb P} 
\newcommand\vv{\mathsf{v}} 
\newcommand\ed{\mathsf{e}} 
\newcommand\ff{\mathsf{f}} 
\newcommand\cc{\mathsf{c}} 
\newcommand\V{\mathsf{V}} 
\newcommand\E{\mathsf{E}} 
\newcommand\F{\mathsf{F}} 
\newcommand\CC{\mathsf{C}} 
\newcommand\Z{\mathsf{Z}} 
\newcommand\D{\Delta} 
\newcommand\Si{\Sigma} 
\newcommand\si{\sigma} 
\newcommand\Q{\mathsf Q} 
\newcommand\cH{\mathcal H} 
\newcommand\cT{\mathcal T} 
\newcommand\cG{\mathcal G} 
\newcommand\cE{\mathcal E} 
\newcommand\cM{\mathcal M} 
\newcommand\cI{\mathcal{I}} 
\newcommand\cN{\mathcal{N}} 

\newcommand\cO{\mathcal O}
\newcommand\e{\epsilon}
\newcommand\ve{\varepsilon}
\newcommand\bn{\boldsymbol{n}}
\newcommand\tK{\tilde K}
\newcommand\tm{\tilde \mu}
\newcommand\tnu{\tilde \nu}
\newcommand\tn{\tilde n}
\newcommand\bpi{\boldsymbol{\pi}}
\newcommand\zb{\bar z}

\newcommand\ra{\mathrm{a}}
\newcommand\sL{\mathscr L}
\newcommand\cL{\mathcal L}

\newcommand\dd{\mathrm{d}} 
\newcommand\ee{\mathrm{e}} 
\newcommand{\grp}[1]{\mathrm{#1}} 

\title{Global magni$4$icence, or: 4G Networks}
\author[a]{Nikita Nekrasov}
\author[b]{Nicolò Piazzalunga}
\affil[a]{Simons Center for Geometry and Physics, Stony Brook University, Stony Brook NY 11794-3636}
\affil[b]{New high energy theory center, Rutgers University}

\begin{document}
\maketitle

\begin{abstract}
\noindent
The global magnificent four theory is the homological version of a
maximally supersymmetric $(8+1)$-dimensional gauge theory on a Calabi--Yau
fourfold fibered over a circle.  In the case of a toric fourfold we
conjecture the formula for its twisted Witten index.  String-theoretically
we count the BPS states of a system of $D0$-$D2$-$D4$-$D6$-$D8$-branes
on the Calabi--Yau fourfold in the presence of a large Neveu--Schwarz
$B$-field.  Mathematically, we develop the equivariant $K$-theoretic
DT4 theory, by constructing the four-valent vertex with generic plane
partition asymptotics.  Physically, the vertex is a supersymmetric
localization of a non-commutative gauge theory in $8+1$ dimensions.

\end{abstract}

\tableofcontents


\section{Introduction}

The discovery \cite{Nakajima:1994nid} of the beautiful connection between
the topology of the moduli spaces of instantons \cite{Belavin:1975fg}
and representation theory of Kac--Moody algebras prompted the search
for the physical explanation of the connection between the seemingly
distinct structures: four-dimensional gauge dynamics on the one hand,
and algebraic structures of two-dimensional conformal field theory,
on another. In physics, the moduli spaces of instantons show up in the
semi-classical evaluation of gauge theory correlation functions. Their
moduli space topology plays a more prominent role in $(4+1)$-dimensional theory,
where the harmonic differential forms on the moduli space of instantons
represent the internal states of solitonic BPS particles. Gauge theories
beyond four dimensions require ultraviolet completion, which is provided
either by string theory, or, in some cases, by the $(2,0)$ superconformal
theory in six dimensions \cite{Seiberg:1996bd}. Computing index of
Dirac operator or its equivariant version, of instanton moduli spaces,
proved quite beneficial in developing the theory of dualities, both
in field theory and in string theory. A useful tool in this endeavour
is the equivariant localization with respect to the isometries of
four-dimensional space ${\BR}^{4}$. Equivariant integrals, for example, can
be computed exactly as sums of local contributions of the fixed points
(or fixed loci more generally). Unfortunately, this is not immediately
useful in the context of instantons, as the fixed instantons are singular,
i.e.\ they are not found in the moduli space $\mathcal{M}$ but rather
on some compactification $\bar{\mathcal{M}}$. Justifying the use of
compactification, e.g.\ used in \cite{Nakajima:1994nid} led to the
noncommutative deformation of gauge theory \cite{Nekrasov:1998s}. In
background independent formulation, the latter can be viewed as
matrix model or matrix quantum mechanics, with a specific type
of infinite matrices \cite{Seiberg:2000zk}. Within this background
independent matrix approach, the computation of path integral of gauge
theory becomes equivalent to an infinite-dimensional version of a
supersymmetric matrix model. Moreover, using various ideas from string
theory \cite{Douglas:1996sw}, one can generalize these matrix models so
as to describe more non-trivial spatial backgrounds, such as ALE spaces,
conifolds or K3 manifolds \cite{Nekrasov:2002kc}. One can observe, that
the intuition of quantum field theory on curved spacetime, in particular
the cluster decomposition, which is a useful method of computations
in topological field theories, persists in the matrix model approach
\cite{Nekrasov2006LocalizingGT}.

One of the fruitful ideas in mathematics is the program of
complexification \cite{Arnold:Complex}, which prompts an eight-dimensional
generalization of instanton enumeration. An early attempt to
formulate such a problem was done in \cite{Baulieu:1997em} in $8$ and in
\cite{Baulieu:1997nj} in $8+1$ dimensions but, without any understanding
of the geometry of the moduli space of generalized instantons let alone
its compactification, the progress was minimal. A recent advance in this
direction was achieved with the introduction \cite{Nekrasov:2015wsu} of
the extension of the ADHM construction of instanton moduli space, which
led to a generalization \cite{Nekrasov:2017cih} of instanton partition
functions. Enumeration of instantons in $4$ or $4+1$ dimensions leads
to the computation of sums over Young diagrams, or $N$-tuples of Young
diagrams for rank $N$ theories. The corresponding generalization to $8$ or
$8+1$ dimensions is a sum over four-dimensional Young diagrams, or solid
partitions. The rank $N$ theories, studied in \cite{Nekrasov:2018xsb}
on ${\BR}^{8} \times S^1$, reduce to the summations of $N$-tuples of
finite size solid partitions.

The goal of this paper is to analyze the theory \cite{Nekrasov:2018xsb}
in the more global setting. We would like to study the super-Yang--Mills
theory in maximal number of space-time dimensions. Classically,
super-Yang--Mills theory can be defined in ten dimensions, with sixteen
real supercharges generating the corresponding supersymmetry. However,
quantum mechanically super-Yang--Mills in ten dimensions suffers from
anomalies, and string theory completion in ten dimensions is only
possible for a very limited class of gauge groups. In nine dimensions
one can use $D8$-branes of $IIA$ string theory to engineer more general
theories. We would like to study the states of the corresponding theory
when $D8$ branes wrap a Calabi--Yau fourfold. The approach taken in this
paper follows the combinatorics \cite{Maulik:2003rzb, Iqbal:2003ds,
Maulik:2004txy} of equivariant Donaldson--Thomas ($DT3$) and equivariant
K-theoretic Donaldson--Thomas theory ($kDT3$), representing
$(6+1)$-dimensional partially twisted maximally supersymmetric Yang--Mills theory
on complex threefolds fibered over a circle \cite{Nekrasov:2014nea}.

The global versions of the Magnificent Four theory have been recently
studied in numerous interesting publications, see \cite{Kanno:2020ybd}
for a review. Both mathematical and physical communities explore this
terra incognita. The main motivation, from our perspective, is to gain
new evidence for $M$-theory, extend the Gromov--Witten/Donaldson--Thomas
correspondence and theory of K{\"a}hler gravity \cite{Maulik:2003rzb,
Iqbal:2003ds, Maulik:2004txy}. There are of course proper mathematical
motivations. The work \cite{Oh:2020rnj, Borisov:2015vha, Cao:2023lon,
Bojko:2020rfg, Cao:2018rft} on mathematical foundations of the four-dimensional
version of Donaldson--Thomas theory $DT4$, the concrete
proposals for signs \cite{Cao:2019tvv,Monavari:2022rtf} (as we
recall below, the orientation of the moduli spaces $\cM_{\zeta}$ is
a nontrivial issue) are a small selection of recent advances. As in
the lower-dimensional cases, the extension to orbifolds is the first
step in the physical approach \cite{Szabo:2023ixw, Kimura:2019msw,
Fucito:2020bjd, Bonelli:2020gku, Cao:2023gvn}. Another version of the
global Magnificent Four theory is the theory on a union of transversal
or intersecting complex surfaces inside a Calabi--Yau fourfold
\cite{Nekrasov:2016gud}, or a similar arrangement of hypersurfaces
\cite{Pomoni:2021hkn}.

Another physically motivated idea is to view the higher-dimensional
instantons as holomorphic maps (quasimaps)
of complex curves into the moduli spaces of instantons
on spaces of two dimension less, cf.~\cite{Atiyah:1984tk,
Bershadsky:1995vm, Losev:1999tu}. One can also study the four-dimensional
analogues of holomorphic maps: the solutions of Seiberg--Witten equations
describing BPS configurations in four-real-dimensional theory with the
gauge group $\grp{U}(k)$ with the matter fields furnishing the ADHM data for
charge $k$ instantons with gauge group $\grp{U}(N)$.  One can also use the
recent studies of the spaces of holomorphic maps of complex surfaces
to K{\"a}hler manifolds \cite{Chekeres:2021xls}.  In our construction
below we combine all these ingredients.

Let $X$ be a toric Calabi--Yau four-fold (smooth quasi-projective toric
variety), with Kähler form $\omega$ and top holomorphic form $\Omega$.
Let $F$ be the curvature of a connection on a complex vector bundle
$\cE$ over $X$, with prescribed Chern character $\ch (\cE)$, satisfying
\cite{Corrigan:1982th} \begin{equation} \label{eq:cyinst} \begin{aligned}
F^{0,2}_+ \coloneq F^{0,2} + \bar \star (\Omega \wedge F^{0,2}) &= 0, \\ F.\omega
&= 0, \end{aligned} \end{equation} where we split $\wedge^2 (T^* X \otimes
\BC) = \langle \omega \rangle \oplus \wedge^{1,1}_0 \oplus \wedge^{2,0}
\oplus \wedge^{0,2}$.  In the decomposition $\wedge^2 (T^* X) = \wedge_7
\oplus \wedge_{21}$ into $Spin(7)$ irreps, \cref{eq:cyinst} corresponds
to the projection $\pi_7 (F)=0$.  Let $\cM$ be the framed moduli space
of solutions to \cref{eq:cyinst} modulo unitary gauge equivalence.
Actually, we work in \emph{non-commutative} gauge theory,\footnote
{Otherwise, Derrick's theorem would apply.  Here we mean gauge theory
in the sense of Seiberg and Witten, namely we investigate open strings
on $X$, in the limit of large $B$-field.  This is what is expected to
produce the higher derivative regularization of PDEs \cref{eq:cyinst},
with the $B$-field corrections. The $B$-field parameters are denoted by
$\zeta$.} and denote the corresponding moduli space by $\cM_\zeta$.

Neglecting torsion, the central charge of a bound state of $D$-branes
in type $IIA$ string theory near large radius is \begin{equation}
\label{charge} Z =  \int_X \ee^{2\pi \ii (B+\ii\omega)} \ch (\cE)
\, \hat \Gamma_X,\end{equation} where the class $\hat \Gamma_X =
\prod_{i=1}^4 \Gamma(1+\delta_i)$ is built out of Chern roots $\delta_i$ of $TX$
and it provides a square-root of A-roof and Todd classes \cite{Iritani:2023ngp}.  We consider
one $D8$-brane, which is infinitely massive and acts as background, see
refs.~\cite{Maulik:2003rzb, Maulik:2004txy, Iqbal:2003ds, Denef:2007vg,
Nekrasov:2018xsb}, as well as one $\overline{D8}$-brane.  We can also
turn on flat Ramond--Ramond background forms $C_{p+1}$ with one leg along
$S^1$, provided they're compatible with toric symmetries, and weigh a
given instanton configuration by \begin{equation} \label{fuga} u \coloneq \exp
\left[ - \int_{X \times S^1} \left( \frac{\dd s}{g_s} \ee^{2\pi \ii (B+\ii\omega)} +\ii
\sum_{n=0}^4 C_{2n+1}\right) \ch (\cE) \, \hat \Gamma_X \right],\end{equation}
where $g_s$ is the string coupling and $\dd s$ a local coordinate on $S^1$.
This $u$ helps keep track of Chern classes $c_i = c_i (\cE)$.  In the
following, we only turn on $C_1$ and define $-p = \exp \int_{S^1}
\frac{\dd s}{g_s}+ \ii C_1$.

\begin{definition} The Donaldson--Thomas partition function is defined
in equivariant\footnote {Equivariant w.r.t.\ the maximal torus $\grp{U}(1)^3$
of $\grp{SU}(4)$ holonomy of $X$, as well as the mass parameter $\tm$.}
K-theory as a generating sum of integrals over virtual fundamental classes
\begin{equation} \label{dt} \Z \coloneq \sum_{(c_1,c_2,c_3,c_4)} u
\int_{[\cM_\zeta]^\mathrm{vir}} \hat A \, \ch \wedge_{\tm}^\bullet E, \end{equation}
where the matter bundle $E$ is the kernel of Dirac operator coupled to
the gauge bundle and it plays the role of insertions in Gromov--Witten
theory \begin{equation} \wedge_{\tm}^\bullet E \coloneq \sum_{i=0}^{\rk E}
(-{\tm})^{i} \wedge^i (E). \end{equation} \end{definition}

\begin{remark} Physically, $\Z$ is the twisted Witten index of a
supersymmetric gauge theory living on the $D8$-brane, up to an overall
perturbative factor. Equivalently, it is a sum of twisted Witten indices
of supersymmetric quantum mechanical models counting bound states of
$D0$-$D2$-$D4$-$D6$-$D8$-branes on $X$, in the limit of large volume
and large $B$-field.\end{remark}

\begin{remark} \label{holo-eul}
Let $\cO^\mathrm{vir}$ be the virtual structure sheaf on $\cM_\zeta$,
which exist thanks to \textcite{Oh:2020rnj},
and $\cE$ the universal sheaf,\footnote
{We previously denoted by the same letter the restriction
of $\cE$ to $X \times \{m\}$, for some $m \in \cM_\zeta$.}
i.e.\ the sheaf on $\cM_\zeta \times X$,
with $\pi$ the projection to the first factor.
The algebro-geometric counterpart of our $\Z$
is the holomorphic Euler characteristic
\begin{equation}
\chi (\cM_\zeta,
\widehat{
\wedge^\bullet_{\tilde\mu} (\pi_{*}\cE) \otimes \mathcal O^\mathrm{vir}
})
\end{equation}
where hat means twist by the square root of determinant
as in \textcite{Nekrasov:2014nea}.
\end{remark}

The main difficulty is to find an orientation on $\cM_\zeta$.  One has
to provide at least an orientation at the fixed points, so that the
integral can be defined via equivariant localization.\footnote {The paper
\cite{Cao:2014bca} mentions the relevance of $DT4$ theory to square root
Euler classes.}

Our strategy is to require covariance of every building block, under
the assumption that the sign choice is essentially unique (up to an
overall sign) and therefore if we can produce a square root that is also
covariant, then it must be the correct answer.
This is related to the fact that the sign induced by the orientation choice
is well-defined \cite{Oh:2020rnj}, so that the choice of both a sign
and of a square-root of virtual tangent bundle is essentially canonical
\cite{Monavari:2022rtf}.

\section{Background material} \subsection{Partitions}

A solid partition $K$ is a collection of non-negative integers
$K_{i,j,k}$ labeled by triples of positive integers, obeying inequalities
\begin{equation} K = \{ K_{i,j,k} \mid K_{i,j,k} \geq \max (0, K_{i+1,j,k},
K_{i,j+1,k}, K_{i,j,k+1}) \, \forall (i,j,k) \in \BZ^3_{>0} \}.
\end{equation} Its size is $|K| = \sum_{i,j,k} K_{i,j,k}$.  Equivalently,
we can represent it by its 4d diagram as \begin{equation} K = \{ (i,j,k,m)
\in \BZ^4_{>0} \mid m \leq K_{i,j,k} \}.
\end{equation} Its character is
\begin{equation} \ch_K (q_1,q_2,q_3,q_4) = \sum_{(k_1,k_2,k_3,k_4) \in
K} \prod_{a=1}^4 q_a^{k_a-1}
\in \BZ [q_1,q_2,q_3,q_4].
\end{equation} In the present setup (local theory) it's
unambiguous to identify a partition with its character.
(Later in \cref{sec-stat}, when we work on a toric variety (global theory),
we must keep track of the variables used to used to compute the character,
which depend on the fixed point as defined in \cref{toric-bg}.)
Given any partition $\rho$, we denote by $\rho^*$ its character evaluated
at the conjugated variables $q_a^*=q_a^{-1}$.  Solid partitions are
in one-to-one correspondence with finite-codimension monomial ideals
$\cI \subset \BC [z_1,z_2,z_3,z_4]$, \begin{equation} \label{kideal}
K = \{ (k_1,\dots, k_4) \in \BZ^4_{>0} \mid \prod_{a=1}^4 z_a^{k_a -1}
\not \in \cI \} \approx \BC [z_1,z_2,z_3,z_4] / {\cI}, \end{equation}
where the $\approx$ symbol means that the set
$\{ \prod_{a=1}^4 z_a^{k_a-1} \mid (k_1,\dots,k_4) \in K\}$
provides a vector space basis of $\BC[z_1,\dots,z_4]/\cI$.
The codimension of $\cI$ equals the size of $K$.  We define partitions
of infinite size by allowing any monomial ideal in \cref{kideal}.
Their character is a formal power series in $\BZ[[q_1,q_2,q_3,q_4]]$.

A similar construction can be performed in any dimension, in particular
we will use $\BC [q_1,q_2]$ (Young diagrams) and $\BC [q_1,q_2,q_3]$
(plane partitions).

A colored partition (plane partition, solid partition) is a vector of
partitions (plane partitions, solid partitions).

\subsubsection[Taylor resolution]{Taylor resolution\footnote
{This section lies somewhat outside the main line of development of this work.}}

Consider a monomial ideal $\cI \subset R = \BC[x_1,\dots,x_n]$, with
generators $m_1,\dots,m_s$ (it is important that $s$ is finite for
a monomial ideal).  For any ordered $T \subseteq \{1,\dots,s\}$, let
$m_T \coloneq \operatorname{lcm} \{ m_i \mid i \in T \}$, the least common
multiple of a subset of generators.  Consider the simplex $F=(F_t)$,
where $F_t = \bigoplus_{|T|=t} R m_T$, for $s \geq t > 1$, has a
formal basis $\{ e_T \colon |T|=t\}$, and the differential \begin{equation}
\dd (e_T) \coloneq \sum_{T'=T\setminus \{i_k\}} (-1)^k \frac{m_T}{m_{T'}}
e_{T'}, \end{equation}
where $T = \{ i_1, \dots, i_{|T|} \}$.
This gives a free resolution of $\cI$, with all
the good properties except that it may be non-minimal.  For example,
the character is \begin{equation} \ch \cI = \sum_{t=1}^s
(-1)^{t+1} \ch F_t. \end{equation}

\subsection{Characters and regularizations} \label{s:bk}

\begin{definition} For some finite set $S \subset {\BZ}^{p}$, we call a
Laurent polynomial \begin{equation} A(x_1,\dots,x_p) = \sum_{{\vec n}
\in S} \, A_{\vec n} \, x_{1}^{n_{1}} x_{2}^{n_{2}} \cdots x_{p}^{n_{p}}
\end{equation} \emph{movable} if it has no constant term, i.e.\ $A_{0}
= 0$.  (We call $A_0$ its \emph{unmovable}, a.k.a.\ fixed, part.)  Let us define the
\emph{mobility} of $A$ by $\# A = \lim_{\boldsymbol x \to 1} (A - A_0)$.
\end{definition}

\begin{definition} The size of a Laurent polynomial $A \in \BZ[x_1^{\pm1},\dots,x_p^{\pm1}]$ is
\begin{equation} |A| = \lim_{\text{all } x \to 1} A, \end{equation}
which can be negative. \end{definition}

\begin{remark}
    Clearly we have $|A| = \# A + A_0$,
    as well as $\#A = \sum_{\vec n \neq 0} A_{\vec n}$.
\end{remark}

\begin{definition} We call ${\chi} \in {\BZ} [[q_1,q_2,q_3,q_4]]$ a
\emph{pure character} if ${\chi} \in \BZ_{\geq 0}[[q_1,q_2,q_3,q_4]]$.
\end{definition}

For $A \subseteq \{ 1,2,3,4 \}$, let $P_A=\prod_{a \in A} (1-q_a)$.

Let $\tK$ be a (possibly infinite) solid partition, with finite
\begin{equation} \tn_{abc} = \lim_{q_a \to 1, q_b \to 1, q_c
\to 1} P_{abc} \tK, \quad a < b < c.\end{equation} Let $\tn =
(\tn_1,\dots,\tn_4) \coloneq (\tn_{234},\dots,\tn_{123})$ and $q^{\tn} \coloneq
q_1 ^{\tn_{1}} q_2 ^{\tn_{2}} q_3 ^{\tn_{3}} q_4 ^{\tn_{4}}$, where we
identify the partition with character $\tn_a (q_a) = 1 + q_a + \dots +
q_a^{\tn_a-1}$ with its size $\tn_a$.

Change variables to \begin{equation} K = q^{-\tn}  \left( \tK -
\frac {1-q^{\tn}}{P_{1234}} \right) \label{KKt} \end{equation} so that
\begin{equation} \lim_{q_a, q_b, q_c \to 1} P_{abc} K = 0, \quad a<b<c.
\end{equation} Its asymptotics determine partitions \begin{equation}
\pi_a (q_1,\dots,\hat q_a,\dots, q_4) = \lim_{q_a \to 1} P_a
K(q_1,\dots,q_4), \end{equation} where hat means removal, and for $a
\neq b$ \begin{equation} \lambda_{ab} (q_1,\dots,\hat q_a,\dots,\hat
q_b,\dots, q_4) =\lim_{q_a, q_b \to 1} P_{ab} K(q_1,\dots,q_4).
\end{equation} Denote $\bpi = (\pi_1,\dots,\pi_4)$.

\begin{remark} Equivalently, if $K$ is defined by a monomial ideal
$\cI$, then \begin{equation} \pi_a = \{ (k_1,\dots, \hat k_a ,\dots,
k_4) \in \BZ^3_{>0} \mid \prod_{b=1}^4 q_b^{k_b -1} \not \in \cI \,
\forall k_a \in \BZ_{>0} \}. \end{equation} In identifying the sets $\pi_a$
and $\lambda_{ab}$ with their characters, it is crucial to keep track
of variables and their ordering correctly: \begin{equation} \ch \pi_a
(q_1,\dots,\hat q_a,\dots, q_4) = \sum_{(k_1,\dots, \hat k_a ,\dots,
k_4) \in \pi_a} \prod_{b \neq a} q_b^{k_b}. \end{equation} \end{remark}

\begin{definition} Introduce regularized partitions \begin{equation}
\label{Kreg}
K_\mathrm{reg} = K - \sum_a \frac{\pi_a}{P_a} + \sum_{a<b}
\frac{\lambda_{ab}}{P_{ab}} \end{equation} and similarly \begin{equation}
\label{pireg}
\pi_{a,\mathrm{reg}} = \pi_a - \sum_{b \neq a} \frac {\lambda_{ab}}{P_b}.
\end{equation} \end{definition}

\begin{remark} We have \begin{equation} K = K_\mathrm{reg} + \sum_a
\frac{\pi_{a,\mathrm{reg}}}{P_a} + \sum_{a<b} \frac{\lambda_{ab}}{P_{ab}}.
\end{equation} \end{remark}

\begin{lemma} The regularized objects are Laurent polynomials.
For example, \begin{equation} \lim_{q_i \to 1} (1-q_i) K_\mathrm{reg}
= 0. \end{equation} \end{lemma}

\subsection{Plethystics}

\begin{definition} The operator $\hat a$ maps a movable Laurent
polynomial $\mathscr P \in \BZ[x_1^\pm,\dots,x_n^\pm]$ to a rational
function $\hat a(\mathscr P)\in\BZ(x_1^{1/2},\ldots,x_n^{1/2})$ such
that \begin{itemize} \item for any monomial $X =
x_1^{c_1} \cdots x_n^{c_n}$ and $p \in \BZ$, $\hat a (p X)=( X^{1/2}
- X^{-1/2} )^{-p}$, \item $\hat a (p X_1 + q X_2) = \hat a(pX_1) \cdot \hat
a (qX_2)$ on monomials $X_1,X_2$ and integers $p,q$. \end{itemize}
\end{definition}

\begin{remark} Such $\hat a$ is the representative in
localization of the A-roof genus.  We will apply it to the ring
$\BZ[q_1^\pm,q_2^\pm,q_3^\pm,\tm^\pm]$.  The definition  involves a
choice of square roots, which is crucial and we postpone until later,
but is unambiguous when operating on perfect squares, namely $\hat a (x +
x^*)$ is independent of such choice.  Note the relation to the plethystic
exponent, $\hat a(x^{-1}) = x^{1/2} \exp \sum_{n=1}^\infty \frac{x^n}{n}$.
\end{remark}

\subsection{Toric geometry}
\label{toric-bg}

Denote the coordinates of $\BC^{N+4}$ by $Z_A$, for $A=1,\ldots, N+4$,
and let the index $i$ run from 1 to $N$.  Let $\Q = (\Q_i^A)$ be an
$N$ by $(N+4)$ matrix of integers.  This defines a $\grp{U}(1)^N$ action on
$\BC^{N+4}$ (for real parameters $\alpha^i$) \begin{equation} Z_A \mapsto
\exp \left( \sqrt{-1} \sum_{i=1}^N  \alpha^i \Q^A_i \right) Z_A \qquad
(A=1,\ldots, N+4). \end{equation} With $m_A = |Z_A|^2$, the momentum map
$\mu: \BR^{N+4} \to \BR^N = (\Lie \grp{U}(1)^N)^\star$ is \begin{equation}
\label{mm} \mu_i (m) \coloneq \sum_{A=1}^{N+4} \Q^A_i m_A, \qquad (i=1,\dots,
N). \end{equation} Let the $N$-tuple of real numbers $r$ be a regular
value of $\mu$.  The quotient $X_r = \mu^{-1}(r)/\grp{U}(1)^N$ is a complex
four-manifold with $\dim H_2(X_r,\BZ)=N$.  It is a Calabi--Yau manifold in
a weak sense\footnote {Strictly speaking, a Calabi--Yau manifold $X$ is a
simply-connected compact Kähler manifold $X$ with vanishing $c_{1}(X)$.}
if \begin{equation} \label{cyc} \sum_{A=1}^{N+4} \Q^A_i = 0 \quad \text{for all
} i=1,\dots,N. \end{equation} In this case it is convenient to define a
map $p \colon \BR^{N+4} \to \BR^{N+3}=(\Lie \grp{U}(1)^{N+3})^\star$ \begin{equation}
p(m)=(m_1-m_2,m_2-m_3,\dots,m_{N+3}-m_{N+4}). \end{equation} Then $\mu$
descends to a map $\BR^{N+3} \to \BR^{N}=(\Lie \grp{U}(1)^{N})^\star$ and the
Newton polyhedron is $\D (X_r) = \mu^{-1}(r) \cap \im (p \circ m)$.
We call its zero-dimensional faces \emph{vertices}, and write $\vv
\in \D_0$.  The vertices have valence four, namely each has four legs
attached to it (some of which may be non-compact), and they correspond to
setting to zero the maximal number of $m_A$'s compatible with \cref{mm}.
Restricting to the compact skeleta of $\D (X_r)$ and regarding them
as oriented, we call the one-dimensional faces \emph{edges}, $\ed
\in \D_1$.  Every edge $\BP^1$ is incident to two vertices.  We call
two-dimensional faces \emph{faces}, $\ff \in \D_2$; and three-dimensional
faces \emph{cells}, $\cc \in \D_3$.  Denote by $|\D_0| = \chi(X_r)$ the
number of vertices, and similarly by $|\D_i|$, for $i=1,2,3$
the number of edges, faces and cells, respectively.

One associates complex line bundles $\cL_i$ ($i=1,\dots,N+4$)
to the $\grp{U}(1)^N$ action and the cohomology of $X$ is generated by
$c_1(\cL_i)$.  The Kähler form is inherited from the ambient vector
space \begin{equation} \omega = \frac{\ii}{2\pi} \sum_A \dd Z_A \wedge \dd \bar
Z_A. \end{equation} The Hamiltonian  is \begin{equation} \label{ham}
H = \sum_{A=1}^{N+4} \ve_A m_A, \end{equation} where $\ve_A$ satisfy
$\sum_A \ve_A =0$ and parametrize the $\grp{U}(1)^{N+3}$ action on $\BC^{N+4}$
\begin{equation} Z_A \mapsto \ee^{\ii \ve_A} Z_A. \end{equation}

At each vertex, we choose four local coordinates $z_a$ built out of
$\grp{U}(1)^N$-invariant combinations of the $Z_A$'s, such that their product
$z_1 z_2 z_3 z_4$ is invariant under $\grp{U}(1)^{N+3}$.  Two vertices connected
by an edge along direction $a$ are related by $z_a \to z_a^{-1}$,
while $z_b \to z_b z_a ^{d_b}$ for $b \neq a$, for some integers $d_b$,
with $\sum_{b \neq a} d_b = 2$ to preserve the Calabi--Yau condition.
The ordering of local coordinates at $\vv$ is inherited from the
orientation of $\D$.  The quotient $\grp{U}(1)^{N+3}/\grp{U}(1)^N$ acts on these four
coordinates, giving local $\Omega$-background parameters $q_a = \ee^{\beta
\e_a}$, with $\sum_{a=1}^4 \e_a =0$.  Sometimes we write $q_a^\vv$ to
emphasize the local dependence.  Two vertices connected by an edge along
direction $a$ are related by $\e_a \to - \e_a$, while $\e_b \to \e_b +
d_b \e_a$ for $b \neq a$.

Denote by $C_i\in H_2(X,\BZ)$ the two-cycles of $X$ with volume
$\int_{C_i} \omega = r_i$.  Dual to them are the (not necessarily compact)
divisors $D^i$, such that $D^j \cdot C_i = \delta^j_i$.  We can think of
the $D^i$ as generating the Kähler cone in $H^2(X,\BZ)$, which equals
by Poincaré duality $H^6_\mathrm{cmpct} (X,\BZ)$.  Denote by $D^A$
the toric divisors $X\cap\{ Z_A=0 \}$, which satisfy $\Q^A_i = C_i
\cdot D^A$.  The Calabi--Yau condition implies trivial canonical bundle,
namely $-\sum_A D^A = 0$, which implies $\sum_A \Q_i^A = 0$ for all $i$.
Alternatively, we have $c(TX) = \prod_A (1+D^A)$.  The compact divisors
correspond to $H_6(X,\BZ)$, which is dual to $H^6(X,\BZ)$ and Poincaré
dual to $H^2_\mathrm{cmpct}(X,\BZ)$.

Since there's a bijection between toric varieties up to isomorphism and
fans up to $\grp{SL}(4,\BZ)$ action, one can also work in the dual picture.
(However, the polyhedron $\D$ includes more data than the dual fan.)
The zero-force condition at each vertex allows to associate to it
a tetrahedron.  Such tetrahedra triangulate the dual toric diagram
(they must have minimal volume $1/3!$ for $X$ to be non-singular).
The vectors $v_i$ that generate the one dimensional cones of the fan
satisfy $\sum_A v_A \Q^A_i = 0$ (in fact, given the fan, one can recover
$\Q$ as the kernel of the matrix of $v_A$'s).

An edge $\ed$ along direction $e$, by Duistermaat--Heckman theorem
has volume \begin{equation} t_\ed = \sum_{\vv \in \ed} \frac{H_\vv}
{\e_e^\vv}, \end{equation}
where $H_\vv$ is the Hamiltonian \cref{ham} restricted at vertex $\vv$.

For a face $\ff$ with tangent directions
$a$,$b$ and normal directions $c$,$d$ define \begin{equation} A_{pqr}
\coloneq \sum_{\vv \in \ff} \frac{(\e_c^\vv)^p (\e_d^\vv)^q H^r_\vv}{\e_a^\vv
\e_b^\vv}, \label{a-funct} \end{equation} where the non-negative integers
$p,q,r$ satisfy $p+q+r=2$ and the sum runs over vertices in the face.
Similarly, define \begin{equation} \label{c2f} c_{2,\ff} \coloneq \sum_{\vv
\in \ff} \frac {c_2} {\e_a \e_b}, \end{equation} where $c_2 = \sum_{1\leq
a<b\leq 4} \e_a \e_b$.

\begin{lemma} Let $\ff$ be a compact face.  Then $A_{110}$ is
even. \label{A110-even} \end{lemma} \begin{proof} Let $S$ be the toric
surface corresponding to $\ff$.
Since $X$ is toric, we can always split $\cN_{S/X} = \sL_3 \oplus \sL_4$,
with \begin{equation} c_1(\sL_3) + c_1(\sL_4)
+ c_1(S) = 0. \end{equation}
The line bundles $\sL_{3,4}$ realizing the splitting of the normal bundle
are closely related to the combinatorial data of the polytope  of $X$,
which is the reason why the normal bundle is split in the first place.\footnote
{We thank the anonymous Referee~2 for pointing this out.}
Then we have $A_{110} = - (c_1^2(\sL_3) +
c_1(\sL_3)c_1(S))$.  Taking $c_1(S)$ as the characteristic element for
the intersection form, we have that $x^2 = c_1(S)x \bmod{2}$ for any $x
\in H^2(S,\BZ)$. \end{proof}

\begin{lemma} \label{b2p1} For any compact face, denoting tangent
directions 1 and 2, we have \begin{equation} \sum_{\vv \in \ff}
\frac{1}{P^*_{12}} = 1. \end{equation} \end{lemma} \begin{proof} For any
projective toric variety $X$, we have $H^m (X,\mathcal O_X)=0$ if $m>0$.
The holomorphic Euler characteristic in the equivariant case is the
alternating sum of the characters, which in the simply connected case
becomes 1 (the character of the global holomorphic functions, which is
one-dim space of constants).  \end{proof}

\begin{remark} The cohomological limit of \cref{b2p1} gives $\frac{\chi +
\sigma}4 = \frac{1+b_{2+}}2 = 1$, which implies $b_{2+}=1$ for projective
toric surfaces.  \end{remark}

\begin{remark} In this notation, the signature of a compact face is
\begin{equation} \sigma = \frac13 \sum_{\vv \in \ff} \frac{(\e_a^\vv)^2+
(\e_b^\vv)^2}{\e_a^\vv \e_b^\vv}. \end{equation} \end{remark}

\section{Statement of the problem}
\label{sec-stat}

Let's focus on the $\grp{U}(1)$ theory, i.e.\ we study a single $D8$-brane.
The fixed points of the torus action on $\cM_\zeta$ are labeled
by collections of solid partitions, possibly of infinite size, $\{
\tK^{\vv}, \ \vv \in \D_0 \}$, satisfying compatibility conditions.
Compatibility means that different $\tK$'s can be glued together
along edges.  For example, suppose $\ed \in \D_1$ joins $\vv_1,
\vv_2\in \D_0$ along the first direction, with local equivariant
parameters $(q_1^{\vv_i},q_2^{\vv_i},q_3^{\vv_i},q_4^{\vv_i})$
for $i=1,2$ related in the standard way.  Then compatibility
requires that if $\lim_{q_1^{\vv_1} \to 1} (1-q_1^{\vv_1})
\tK^{\vv_1} (q_1^{\vv_1},q_2^{\vv_1},q_3^{\vv_1},q_4^{\vv_1})=
\pi^{(\ed)} (q_2^{\vv_1},q_3^{\vv_1},q_4^{\vv_1})$,
then $\lim_{q_1^{\vv_2} \to 1} (1-q_1^{\vv_2}) \tK^{\vv_2}
(q_1^{\vv_2},q_2^{\vv_2},q_3^{\vv_2},q_4^{\vv_2})= \pi^{(\ed)}
(q_2^{\vv_2},q_3^{\vv_2},q_4^{\vv_2})$.  A similar equation holds for
double limits and $\lambda$'s along faces.

\begin{definition}
    Let $\tnu$ be the character of the framing space,
    i.e.\ the Coulomb branch parameter of our $\grp{U}(1)$ theory.
\end{definition}
\begin{remark} Without loss of generality, we can set $\tnu=1$ in the
$\grp{U}(1)$ theory.  \end{remark}

\begin{definition} \label{univ} The basic object of the local theory
is \begin{equation} \cH \coloneq \frac{\tnu - \tm}{P_{1234}} - \tnu
\tK = q^{\tn} \left( \frac{\tnu - \mu}{P_{1234}} - \tnu K \right),
\end{equation} where $\tm$ is part of definition \cref{dt},
$\mu=q^{-\tn} \tilde \mu$ and\footnote
{We are using the same letter to denote
the equivariant mass parameter $\mu^{(\vv)}$ for $\vv \in \Delta_0$
and the vector of moment maps $\mu_i (\cdot)$, for $i =1,\ldots,N$ in \cref{mm}.}
we used \cref{KKt}.
Define the virtual character \begin{equation} \label{t2} T^2 [K] \coloneq
- P_{1234} \cH \cH^* - T^2_\mathrm{pert}(\mu), \end{equation} where
we subtracted the perturbative part \begin{equation} \label{t2pert}
T^2_\mathrm{pert}(\mu) \coloneq \frac{ (\tnu-\mu) (\tnu-\mu)^*}{P^*_{1234}}.
\end{equation} This corresponds to an infinite-dimensional space, as it
has poles.  We can write the perturbative part as the sum of a finite
contribution (depending on $\tn$) \begin{equation} \label{ts6} T_6^2 \coloneq
\sum_{\vv\in \D_0} (T^2_\mathrm{pert}(\mu) - T^2_\mathrm{pert}(\tm))
\end{equation} and an infinite one (independent of $\tn$).  Around
$\vv \in \D_0$, we associate to $\ff \in \D_2$ along directions $a,b$
the character \begin{equation} \label{face2} \cT ^2_\ff \coloneq T^2 \left[
\frac{\lambda_{ab}}{P_{ab}} \right]. \end{equation} To $\ed\in \D_1$
along direction $e$, we associate $\cT^2_\ed$, defined by the formula
\begin{equation} \label{tted} T^2 \left[ \frac{\pi_e}{P_e} \right] = \cT^2_\ed +
\sum_{\ff \mid \ed \in \ff} \cT^2_\ff. \end{equation} Finally, we define
$T^2_{\vv}$ such that \begin{equation} \label{T2K} T^2[K] = T^2_\vv + \sum_{\ed \mid
\vv\in\ed} \cT^2_\ed + \sum_{\ff \mid \vv\in\ff} \cT^2_\ff. \end{equation}
\end{definition}

\begin{remark} The perturbative part $T^2_\mathrm{pert}(\mu)$ depends on $\tn$
through $\mu$.  However, $\tn$ does not couple to lower-dimensional
partitions.  \end{remark}

\begin{remark} Mutatis mutandis, the same structure appears in complex
dimensions $d=2,3,4$ \begin{equation} T [K] = -P_{1 \dots d} \cH \cH^*
- T_\mathrm{pert} \end{equation} with contributions from codimension-one
objects $\cc \in \D_{d-1}$ (almost) tensored away as in \cref{KKt}.
However, due to the reality of \cref{eq:cyinst} (real representation
of $\grp{SU}(4)$), in $d=4$ for the first time we have to take half of the
equations.  \end{remark}

\begin{lemma} A computation shows that \begin{multline} \label{edge2}
\cT^2_\ed = \left( 1 - \mu - P_{1234} \sum_{ a \neq e} \frac{
\lambda_{ae}}{P_{ae}} \right) \frac{\pi^*_{e,\mathrm{reg}}}{P^*_e} +Q
\left( 1 - \mu - P_{1234} \sum_{ a \neq e} \frac{ \lambda_{ae}}{P_{ae}}
\right)^* \frac{\pi_{e,\mathrm{reg}}}{P_e} \\ - P_{1234} \frac
{\pi_{e,\mathrm{reg}}}{P_e} \frac{\pi^*_{e,\mathrm{reg}}}{P^*_e} -
P_{1234} \sum_{a \neq e, b \neq e, a \neq b} \frac{\lambda_{ae}}{P_{ae}}
\frac{\lambda^*_{be}}{P^*_{be}}, \end{multline} where $Q=\prod_{a=1}^4
q_a$, and, if we denote by $\sum'$ the sum over pairwise distinct
$a,b,c,d$, \begin{multline} \label{vtx2} T^2_\vv = \left(1 - \mu
-P_{1234} \left(\sum_a \frac{\pi_{a,\mathrm{reg}}}{P_a} + \sum_{a < b}
\frac{\lambda_{ab}}{P_{ab}} \right)\right) K^*_\mathrm{reg} \\ +Q\left(1 -
\mu -P_{1234} \left(\sum_a \frac{\pi_{a,\mathrm{reg}}}{P_a} + \sum_{a <
b} \frac{\lambda_{ab}}{P_{ab}} \right)\right)^* K_\mathrm{reg} -P_{1234}
K_\mathrm{reg} K^*_\mathrm{reg} \\ -P_{1234}  \left( \sum_{a \neq b}
\frac{\pi_{a,\mathrm{reg}}}{P_a} \frac{\pi^*_{b,\mathrm{reg}}}{P_b^*}
+ \left( \sum_a \frac{\pi_{a,\mathrm{reg}}}{P_a} \sum_{c<d;c,d \neq a}
\frac{\lambda^*_{cd}}{P^*_{cd}} + \text{c.c.} \right) + \sum'_{a<b,c<d}
\frac{\lambda_{ab}}{P_{ab}} \frac{\lambda^*_{cd}}{P^*_{cd}} \right).
\end{multline}
Here $\text{c.c.}$ means the complex conjugate.
\end{lemma}
\begin{proof}
The first part is obtained by plugging \cref{face2,t2,t2pert} into \cref{tted},
and using \cref{pireg} as well as the definition of $Q$ in the lemma.
Observe that $P_{1234}^* Q = P_{1234}$.

Substituting the first part into \cref{T2K}, and using \cref{Kreg,pireg},
which imply that
$K = K_\mathrm{reg} + \sum_a \frac{\pi_{a,\mathrm{reg}}}{P_a}
+ \sum_{a<b} \frac{\lambda_{ab}}{P_{ab}}$,
after a similar computation we get the second part.
\end{proof}

Globally, each $T^2_\vv$ depends on local coordinates $q_a^\vv$ around
$\vv \in \D_0$ (also possibly through $\mu^{(\vv)}$); similarly, each
summand in $T^2_\ed := \sum_{\vv \in \ed} \cT_\ed^2$ (two summands) and
$T_\ff ^2 := \sum_ {\vv \in \ff} \cT_\ff^2$ (arbitrary number $\geq 3$
of summands) depends on its local coordinates.  The redistribution
\begin{equation} \label{glue} \begin{aligned} \sum_{\vv \in \D_0}
T^2 [K^\vv ] &= \sum_{\vv \in \D_0} T^2_\vv + \sum_{\vv \in \D_0}
\sum_{\ed \mid \vv} \cT_\ed ^2 + \sum_{\vv \in \D_0} \sum_{\ff \mid \vv}
\cT_\ff^2 \\ & = \sum_{\vv \in \D_0} T^2_\vv + \sum_{\ed \in \D_1}
\sum_{\vv \in \ed} \cT_\ed^2 + \sum_{\ff \in \D_2} \sum_{\vv \in \ff}
\cT_\ff^2 \\ & = \sum_{\vv \in \D_0} T^2_\vv + \sum_{\ed \in \D_1}
T_\ed^2 + \sum_{\ff \in \D_2} T_\ff^2 \end{aligned} \end{equation}
is such that $T^2_\vv$, $T^2_\ed$, $T_\ff ^2$ (as well as $T^2_6$) are
movable Laurent polynomials, as we discuss below.  The
equivariant K-theory class of the
virtual tangent space to $\cM_\zeta$ at a fixed point (including contributions
from the matter bundle $E$) is the virtual character \begin{equation} I^2
\coloneq T^2_6 + \sum_{\vv \in \D_0} T^2_\vv + \sum_{\ed \in \D_1} T_\ed^2 +
\sum_{\ff \in \D_2} T_\ff^2. \end{equation}

Our goal is to apply $\hat a$ to $I^2$, after taking a suitable square
root.  All these terms are squares in the sense that if they contain
a monomial $m$, then they also contain $m^*$, and taking square roots
means consistently picking only half of these terms.
The fact that $I^2$ is a square has a geometric interpretation
in the deformation-obstruction complex of $X$ being self-dual.

\begin{remark} \label{univ-inst}
Consider the pullback $\cE_\vv = \iota_\vv^* \cE$ of the universal sheaf
via the inclusion $\iota_\vv \colon \cM \times \{\vv\} \to \cM \times X$
of a fixed point $\vv \in \Delta_0$.  With $N=q^{\tn}\tnu$, we have
\begin{equation} \ch \cE_\vv = N - P_{1234} K.
\end{equation}
Then, we get
\begin{multline}
    \int_X \left[ \ch(\cE\otimes\cE^*) - \ch(\cE\otimes M^*)
    - \ch(\cE^*\otimes M) \right]
    \td_X \\ = 
    \sum_{\vv\in\D_0}
    \frac{\ch\cE_\vv\otimes\cE_\vv^*
    -\ch\cE_\vv\otimes M^*
    -\ch\cE^*_\vv\otimes M}{P_{1234}^*} \\
    = \sum_{\vv\in\D_0} P_{1234}\cH\cH^*,
\end{multline}
the last equality up to an irrelevant (non-dynamical) factor $MM^*$.
The gauge theoretic representative of $\ch\cE_\vv$ also gives rise
to the infinite factors $NN^*/P_{1234}$, the perturbative part,
which we subtract accordingly.  The result equals $\sum_{\vv\in\D_0} T^2[K^\vv]$.
Finally, the map $\hat a$ converts this from the equivariant
K-theory of the fixed locus in $\cM$ to its localization.
\end{remark}

\section{Computation of fugacities}

Each vertex $\vv \in \Delta_0$ sits at the intersection of four toric
divisors, some of which can be non-compact.  Compact toric divisors of
$X$ correspond to cells, and to each cell $\cc \in \Delta_3$
we associate an integer $\tn_\cc$.  Let $S_\vv \subset \{1,2,3,4\}$
be the set of directions that are normal to compact divisors at $\vv$,
and define a map $i_\vv \colon S_\vv \to \Delta_3$ that assigns to each normal
direction its cell, so that we can write $\e \cdot \tn = \sum_{a \in
S_\vv} \e_a^{(\vv)} \tn_{i_\vv (a)}$.  Equivalently, we have a line
bundle $L$ with equivariant first Chern class $c_1(L) = \sum_{\cc \in
\Delta_3} \tn_\cc c_1 (D_\cc)$, where divisor $D_\cc$ corresponds
to cell $\cc$.

As in \cref{univ-inst},
consider the pullback $\cE_\vv = \iota_\vv^* \cE$ of the universal sheaf
via the inclusion $\iota_\vv \colon \cM \times \{\vv\} \to \cM \times X$
of a fixed point $\vv \in \Delta_0$.
With $\tnu = \ee^{\beta \tilde \alpha}$ and $N=q^{\tn}\tnu$, its character
\begin{equation} \ch \cE_\vv = N - P_{1234} K = q^{\tn}\tnu -
q^{\tn}\tnu P_{1234} \left( K_\mathrm{reg} + \sum_a \frac{\pi_{a,\mathrm{reg}}}{P_a} +
\sum_{a<b} \frac{\lambda_{ab}}{P_{ab}} \right) \end{equation} can be
conveniently written as the tensor product $\ch \cE_\vv = \ch (L \otimes
\cE_\vv^{(0)})$, where $\cE_\vv^{(0)}$ is simply $\cE_\vv$ with all the fluxes $\tn$
set to zero, and $\ch L = q^{\tn}=\ee^{\beta \tn\cdot\e}$.

The equivariant gamma-class of a toric CY 4-fold at a fixed point
$\vv \in \Delta_0$ \begin{equation} \hat \Gamma_\vv = \prod_{a=1}^4 \Gamma
\left(1 + \frac{\beta}{2\pi} \e_a\right) = 1 - \frac{1}{24} c_2 \beta^2
- \frac{\zeta(3)}{(2\pi)^3} c_3 \beta^3 + \frac{\zeta(4)}{(2\pi)^4}
\left(\frac74 c_2^2 - c_4 \right) \beta^4 + O(\beta^5) \end{equation}
can be written in terms of elementary symmetric polynomials $c_i$ in
variables $\e_1,\dots,\e_4$.

We want to use equivariant localization to compute \cref{charge} in
terms of \begin{equation} \label{Z2n}
 Z_{2n} \coloneq \int_X \frac{(-\omega)^n}{n!} (\ch (\cE)\,\hat \Gamma_X)_{4-n}
 = \sum_{\vv \in \D_0} \frac{H^n_\vv}{n!} \frac1{\prod_{a=1}^4 \e_a^\vv}
 (\ch\cE_\vv\cdot\hat\Gamma_\vv)_{4-n}, \quad 0 \leq n \leq 4
\end{equation} where the subscript $4-n$ denotes the power of $\beta$,
and $H_\vv$ is the Hamiltonian evaluated at $\vv$.  Recall that $Z_0$
appears as $(-p)^{-Z_0}$ in the fugacity $u$.

\begin{definition} For any edge $\ed \in \Delta_1$, we have transition
functions $d_i$ defined in \cref{toric-bg} obeying the relation $\sum_i
d_i = 2$.  Given a (regularized) plane partition $\pi$, its box is denoted
as $\Box=(i-1,j-1,k-1)$, and we introduce the sign $s_\pi (\Box)=\pm1$,
depending on whether the box is added or removed (just like the size of
a regularized partition can be negative).  Let \begin{equation} f_\ed
(\pi) \coloneq \sum_{\Box \in \pi} s_\pi (\Box) [ 1 - i d_i - j d_j - k d_k ].
\end{equation} Given a tuple of integers $(\tn_\cc)_{\cc\in\Delta_3}$,
let\footnote{Here and below, we are denoting by the same letter
different functions, which are distinguished only by the argument
they take.} \begin{equation} f_\ed (\tn) \coloneq \sum_{\vv \in \ed} \frac
{\e\cdot\tn}{\e_\ed}, \end{equation} where the sum runs over the two
vertices in the edge, and $\e_\ed$ denotes tangent direction.

For any face $\ff\in\Delta_2$, define functions $A_{pqr}$ as in
\cref{a-funct}.  Given a Young diagram $\lambda$, denote its box by $\Box
= (i-1,j-1)$, and let \begin{equation} g_\ff (\lambda) \coloneq \sum_{\Box
\in \lambda} \left[ \left(\frac{i(i-1)}2+\frac16\right) A_{200}+
\left(\frac{j(j-1)}2+\frac16\right) A_{020} + \left(ij-\frac{i+j}2
+ \frac14 \right) A_{110} \right]. \end{equation} Similarly, let
\begin{equation} \tilde g_\ff (\lambda) \coloneq \sum_{\Box \in \lambda}
\left[ (i-1/2) A_{101} + (j-1/2) A_{011} \right]. \end{equation} Given
a tuple of integers $(\tn_\cc)_{\cc\in\Delta_3}$, let \begin{equation}
g_\ff (\lambda,\tn) := \sum_{\vv \in \ff} \frac {(i-1/2) \e_c + (j-1/2)
\e_d} {\e_a \e_b} \, (\e \cdot \tn), \end{equation} where $a$, $b$ denote
directions tangent to $\ff$, $c$, $d$ denote directions normal to $\ff$,
and the sum runs over vertices in $\ff$.  Similarly, let \begin{equation}
\tilde g_\ff (\tn) \coloneq \sum_{\vv \in \ff} \frac {(\e \cdot \tn) \, H_\vv}
{\e_a \e_b} \end{equation} as well as \begin{equation} g_\ff (\tn) \coloneq
\sum_{\vv \in \ff} \frac {(\e \cdot \tn)^2} {\e_a \e_b}. \end{equation}

Finally, define global functions \begin{equation} c_{p,q,r} (\tn) \coloneq
\sum_{\vv \in \Delta_0} \frac {c_p (\tn\cdot\e)^q H^r_\vv} {\prod_{a=1}^4
\e_a}, \quad (p,q,r) = (2,1,1), (2,2,0), (3,1,0) \end{equation} as
well as \begin{equation} h_i (\tn) \coloneq \sum_{\vv \in \Delta_0} \frac
{H_\vv^{4-i}}{(4-i)!} \frac {(\tn \cdot\e)^i}{i!} \frac 1 {\prod_{a=1}^4
\e_a}, \quad i=1,2,3,4. \end{equation}  \end{definition}

\begin{lemma} \label{charges} Terms involving $\tilde\alpha$ do not
contribute to the sum over fixed points in \cref{Z2n}.  Up to terms
independent of $\cE$, which contribute overall factors, the $D0$-brane
charge reads \begin{multline} Z_0 = - \sum_{\vv\in\Delta_0} |K_\mathrm{reg}^\vv|
+ \sum_{\ed\in\Delta_1} f_\ed (\pi_{\ed,\mathrm{reg}}) + \sum_{\ed \in \Delta_1}
|\pi_{\ed,\mathrm{reg}}| f_\ed (\tn) - \sum_{\ff\in\Delta_2} g_\ff (\lambda_\ff)
+ \frac{1}{24} \sum_{\ff \in \Delta_2} |\lambda_\ff| c_{2,\ff} \\ -
\sum_{\ff \in \Delta_2} g_\ff (\lambda_\ff,\tn) - \frac12 \sum_{\ff \in \Delta_2}
|\lambda_\ff| g_\ff (\tn) + h_4 (\tn) - \frac12 \frac{1}{24} c_{2,2,0}
(\tn) - \frac{\zeta(3)}{(2\pi)^3} c_{3,1,0} (\tn) \end{multline}
with $c_{2\ff}$ defined as in \cref{c2f}.  The D2-brane charge
reads \begin{equation}\label{d2} Z_2 = \sum_{\ed\in\Delta_1} t_\ed
|\pi_{\ed,\mathrm{reg}}| - \sum_{\ff\in\Delta_2} \tilde g_\ff (\lambda_\ff) -
\sum_{\ff \in \Delta_2} |\lambda_\ff| \tilde g_\ff (\tn) + h_3 (\tn) -
\frac{1}{24} c_{2,1,1} (\tn). \end{equation} The D4-brane charge reads
\begin{equation} \label{d4} Z_4 = - \frac12 \sum_{\ff \in \Delta_2}
|\lambda_\ff| A_{002} + h_2(\tn). \end{equation} The D6-brane charge reads
\begin{equation} \label{d6} Z_6 = h_1 (\tn). \end{equation} \end{lemma}

\begin{proof}
Let us drop the suffix $\vv$ for brevity, and
expand $\hat\Gamma = 1 + \sum_{i \geq 2} \Gamma_i
\beta^i$ and $\ch L = 1 + \sum_{i \geq 1} L_i\beta^i$, where $L_i =
\frac{1}{i!} (\tn\cdot\e)^i$.  Let us expand $\ch \cE^{(0)} = 1 + \sum_{i
\geq 1} \cE^{(0)}_i \beta^i$ and compute some terms: \begin{multline}
\cE^{(0)}_4  = \frac {\tilde\alpha^4}{4!} - |K_\mathrm{reg}| \prod_{a=1}^4
\e_a + \sum_a \sum_{\Box \in \pi_{a,\mathrm{reg}}} \left( \Box - \frac12 \e_a
+ \tilde\alpha \right) \prod_{b \neq a} \e_b \\ - \frac12 \sum_{a<b}
\sum_{\Box \in \lambda_{ab}} \left[ (\Box + \tilde\alpha)^2 + (\Box +
\tilde\alpha) \sum_{c\neq a,b} \e_c + \frac13 \sum_{c \neq a,b}\e_c^2
+\frac12 \prod_{c \neq a,b} \e_c \right] \prod_{c \neq a,b} \e_c,
\end{multline} \begin{equation} \cE^{(0)}_3  = \frac {\tilde\alpha^3}{3!}
+ \sum_a |\pi_{a,\mathrm{reg}}| \prod_{b \neq a} \e_b - \sum_{a<b} \sum_{\Box \in
\lambda_{ab}} \left( \Box + \tilde\alpha + \frac12 \sum_{c \neq a,b}
\e_c \right) \prod_{c \neq a,b} \e_c, \end{equation} \begin{equation}
\cE^{(0)}_2 = \frac {\tilde\alpha^2}{2} - \sum_{a<b} |\lambda_{ab}|
\prod_{c \neq a,b} \e_c, \end{equation} \begin{equation} \cE^{(0)}_1 =
\tilde\alpha. \end{equation}

The relevant product is then \begin{multline} \ch\cE \cdot \hat\Gamma
= \ch L \cdot \ch \cE^{(0)} \cdot \hat \Gamma = 1 + \beta (L_1 +
\cE^{(0)}_1) + \beta^2 (\cE^{(0)}_2 + \Gamma_2 + L_2 + \cE^{(0)}_1 L_1) \\
+ \beta^3 (\cE^{(0)}_3 + \Gamma_3 + L_3 + \cE^{(0)}_2 L_1 + \cE^{(0)}_1 L_2
+ \cE^{(0)}_1 \Gamma_2 + \Gamma_2 L_1) \\ + \beta^4 (\cE^{(0)}_4 + L_4 +
\Gamma_4 + \cE^{(0)}_3 L_1 + \cE^{(0)}_2 \Gamma_2 + \cE^{(0)}_2 L_2 +
\cE^{(0)}_1 \Gamma_3 + \cE^{(0)}_1 L_3 + L_2 \Gamma_2 + L_1 \Gamma_3).
\end{multline} Summing over fixed points we get the result. \end{proof}

\section{Vertex}

In this section, we introduce the K-theoretic four-vertex.  This is
where most of the complexity of DT counts lies.  First, we introduce
some machinery to deal with partitions of infinite size, then take a
candidate square-root of the vertex, and fix its sign using residues.

\subsection{Euler Characteristic}

\begin{definition} Given a quadruple of plane partitions
$\bpi=(\pi_1,\pi_2,\pi_3,\pi_4),$ possibly of infinite size, introduce
the sets \begin{equation} \begin{aligned} \Si_1 &= \{ (\bn;a) \mid \bn
\in \BZ^4_>, 1 \leq a \leq 4, \bn_a \in \pi_a \}, \\ \Si_2 &= \{ (\bn;a,b)
\mid \bn \in \BZ^4_>, 1 \leq a < b \leq 4, \bn_a \in \pi_a, \bn_b \in
\pi_b \},\\ \Si_3 &= \{ (\bn;a,b,c) \mid \bn \in \BZ^4_>, 1 \leq a < b <
c \leq 4, \bn_a \in \pi_a, \bn_b \in \pi_b, \bn_c \in \pi_c \}, \\ \Si_4 &=
\{ \bn \mid \bn \in \BZ^4_>, \bn_a \in \pi_a \, \forall a=1,\dots,4 \},
\end{aligned} \end{equation} where $\bn_a$ for $1 \leq a \leq 4$ means
$\bn$ with $a$-th entry dropped.  As the $\Si_i$'s were defined as sets,
we define their characters: \begin{equation} \begin{aligned} \ch \Si_1 &=
\sum_{(\bn; a) \in \Si_1} \prod_{i=1}^4 q_i^{(\bn)_i-1}, \\ \ch \Si_2 &=
\sum_{(\bn; a,b) \in \Si_2} \prod_{i=1}^4 q_i^{(\bn)_i-1}, \\ \ch \Si_3 &=
\sum_{(\bn; a,b,c) \in \Si_3} \prod_{i=1}^4 q_i^{(\bn)_i-1}, \\ \ch \Si_4
&= \sum_{ \bn \in \Si_4} \prod_{i=1}^4 q_i^{(\bn)_i-1}, \end{aligned}
\end{equation} where $(\bn)_i$ denotes the $i$-th component of $\bn$.
We introduce the set $\Si$, defined through its character \begin{equation}
\label{Sigma} \ch \Si \coloneq \ch \Si_1 - \ch \Si_2 + \ch \Si_3 - \ch \Si_4.
\end{equation} Given a solid partition $K$ with asymptotics $\bpi$,
define its `Euler characteristic' \begin{equation} \label{chi} \chi_K \coloneq
\ch K - \ch \Si. \end{equation} \end{definition}

\begin{remark} The character $\Si$ is a pure character and the character
of some infinite partition.  Moreover, $P_{1234} \Si$ is a Laurent
polynomial.  \end{remark}

\begin{remark} The sets $\Si_1$, $\Si_2$, $\Si_3$, $\Si_4$ do not depend
on $K$.  We have $\ch\Si_1 = \sum_a \frac {\pi_a} {P_a}$.  One can
rewrite \begin{equation} \Si_2 = \{ (\bn;a,b) \mid (\bn,a) \in \Si_1,
(\bn,b) \in \Si_1, a<b \}. \end{equation} \end{remark}

\begin{lemma} The character $\chi_K$ is a pure character and a Laurent
polynomial.  \end{lemma} \begin{proof} Suppose $\lim_{q_i \to 1} (1-q_i)
\chi_K = \rho_i$.  Then $\chi_K$ must contain the whole series $S =
\frac{\rho_i}{1-q_i}$ and since $\chi_K \subseteq K$ then $S \subseteq K$.
But then it must be $\rho_i \subseteq \pi_i$, therefore $S \subseteq
\Si_1$.  This is in contradiction with the fact that if $\bn \in \chi_K$
then $\bn \not \in \Si_1$.  \end{proof}

\begin{definition} Given a quadruple of plane partitions
$\bpi=(\pi_1,\pi_2,\pi_3,\pi_4),$ possibly of infinite size, recall
the identities \begin{equation} \lambda_{ab} \coloneq \lim_{q_a \to 1}
(1-q_a) \pi_b = \lim_{q_b \to 1} (1-q_b) \pi_a, \qquad 1 \leq a<b \leq 4
\end{equation} and introduce the set \begin{equation} \Si_2' = \{ (\bn;
a,b) \mid \bn \in \BZ^4_>, 1 \leq a<b \leq 4, \bn_{ab} \in \lambda_{ab} \},
\end{equation} where $\bn_{ab}$ for $1\leq a < b \leq 4$ means $\bn$ with
the $a$-th and $b$-th entries dropped. Its character is \begin{equation}
\ch \Si_2' = \sum_{(\bn; a,b) \in \Si_2'} \prod_{i=1}^4 q_i^{(\bn)_i-1}.
\end{equation} \end{definition}

\begin{remark} The set $\Si_2' \subseteq \Si_2$ does not depend on $K$.
We have $\ch\Si_2' = \sum_{a<b} \frac {\lambda_{ab}} {P_{ab}}$.
\end{remark}

\begin{definition} Define the difference of Laurent polynomials
\begin{equation} \label{ell} \ell \coloneq K_\mathrm{reg} - \chi_K =
\ch \Si - \ch \Si_1 +
\ch \Si_2'. \end{equation} \end{definition}

\begin{remark} The Laurent polynomial $\ell$ only depends on $\bpi$
(not on $K$).  \end{remark}

\begin{remark}  A similar construction of the pair $(\chi,\Sigma)$ can
be performed in any dimension $d$.  For Young diagrams ($d=2$), $\chi$ is
still a Young diagram, while this is not the case in general. \end{remark}

\subsection{Square roots}
Within this section, we fix $\vv \in \D_0$.
\begin{definition}
Let us define the virtual character \begin{equation} \label{tcross}
T_\mathrm{cross} \coloneq - \sum_{a \neq b} P_{d(a,b)} q_b \pi_{a,\mathrm{reg}}
\pi^*_{b,\mathrm{reg}} - P_{1234} \sum_a \frac{\pi_{a,\mathrm{reg}}}{P_a}
\sum_{\substack{c<d\\c,d \neq a}} \frac{\lambda^*_{cd}}{P^*_{cd}}
- \sum_{\substack{(abcd)=(1234),\\(1324),(1423)}} \lambda_{ab}
\lambda^*_{cd} q_c q_d, \end{equation}
where $d(a,b)$
is the smallest integer in $\{1,2,3,4\}$ different from  $a$ and $b$.
This $T_\mathrm{cross}$ contains constant terms (all
manifestly movable) from the viewpoint of the vertex.\footnote{
We call a virtual character \emph{constant} from the viewpoint
of the vertex if it is the same for
any two solid partitions $K_\vv$ and $K'_\vv$
that have the same asymptotics $\bpi$.}  
\end{definition}

\begin{remark} This choice depends on the ordering of the set of edges
emanating from a vertex. When gluing vertices, we will need to check
the independence of the final result.  \end{remark}

\begin{lemma}
After imposing $Q=1$, the virtual character
\begin{equation} \label{vtx} T_\vv = \left( 1 - \mu -P_{1234}
\left( \sum_a \frac{\pi_{a,\mathrm{reg}}}{P_a} + \sum_{ a < b}
\frac{ \lambda_{ab}}{P_{ab}} \right)\right)^* K_\mathrm{reg}
-P_{123} K_\mathrm{reg} K^*_\mathrm{reg} + T_\mathrm{cross}
\end{equation} is a square root of \cref{vtx2}.
Both $T_\vv$ and $T_\mathrm{cross}$ are manifestly
Laurent polynomials.
\end{lemma}
\begin{proof}
Use the identities $P_{abc} + P^*_{abc} = P_{1234}$
and $q_d P_d^*=-P_d$, as well as
$q_a q_c = (q_d q_b)^*$ for any permutation $(a,b,c,d)$ of $(1,2,3,4)$.
\end{proof}

\begin{remark} The condition $Q=1$, once imposed at one vertex, is
satisfied at every vertex.  \end{remark}

\begin{definition}For a given $K$,
with $\chi \coloneq \chi_K$ and $\Si$ defined in \cref{chi,Sigma},
respectively,
let us define the non-constant part \begin{equation} T \coloneq
(1-\mu)^* \chi - P_{123} \chi \chi^* - P_{1234}\Si^* \chi. \label{T}
\end{equation} With $\ell$ as in \cref{ell}, let us define
the constant part \begin{equation} \label{tconst} T_\mathrm{const} \coloneq
(1-\mu-P_{1234} (\Si-\ell))^* \ell - P_{123} \ell \ell^*. \end{equation}
\end{definition}

\begin{lemma}
Up to conjugation,
the Laurent polynomial $T_\vv$ decomposes as
\begin{equation} T_\vv = T + T_\mathrm{cross} + T_\mathrm{const}.
\end{equation} 
\end{lemma}
\begin{proof}
Recall that $K - K_\mathrm{reg} = \sum_a \frac{\pi_{a,\mathrm{reg}}}{P_a}
+ \sum_{a<b} \frac{\lambda_{ab}}{P_{ab}}$.
Plug in the definitions, use the fact that $\ell$ is a Laurent polynomial,
and conjugate the finite term $P_{123}\ell \chi^*$.
\end{proof}

\begin{lemma} The Laurent polynomial $T$ is movable.  \end{lemma}
\begin{proof} Recall that $\Si$ is a pure character and the character of
some (infinite) solid partition.  We use induction on the size of $\chi$.
For $\chi=0$, $T=0$ is movable.  Denote the new box by $\xi$.
We work up to movable terms, i.e.\ $(\cdot)_0$ is understood
everywhere in the proof. We have
\begin{equation} \delta T \coloneq T[\chi+\xi]-T[\chi] = \xi +
1 -P_{123} (\chi+\xi)^*\xi-P_{123} (\chi+\xi)\xi^* -P_{1234}\Sigma^*
\xi. \end{equation} Let $\Pi = \Pi(\xi)$ be the parallelepiped generated
by $\xi$, and $\Pi_1 = (\chi+\xi)\cap \Pi$.  Then $\chi+\xi = \Pi_1 +
(\chi+\xi)\setminus \Pi_1$.  By the same arguments as the finite case
\cite[Section 2.4.1]{Nekrasov:2018xsb}, $P_{123} ( (\chi+\xi) \setminus \Pi_1)^*
\xi=0$ and $P_{123} ((\chi+\xi) \setminus \Pi_1) \xi^*=0$.  Then
\begin{equation} \delta T = 1 + \xi -P_{123} \Pi_1^* \xi -P_{123} \Pi_1
\xi^*-P_{1234}\Sigma^* \xi. \end{equation} Now we claim that $P_{1234}
(\Sigma \setminus \Pi_2)^* \xi=0$, where $\Pi_2 = \Sigma \cap \Pi$.
Then \begin{equation} \delta T = 1+ \xi -P_{1234} (\Pi_1+\Pi_2)^* \xi
\end{equation} but $\Pi_1 + \Pi_2 = \Pi$ and $P_{1234} \Pi^* \xi = 1 +
\xi$ is literally the finite case result.
\end{proof}

\begin{definition} \label{vtx-res}
Let $k=|\chi|$.
By replacing $\chi \to \sum_{i=1}^k x_i$ in $T$,
let \begin{equation} T_\mathrm{formal} = (1-\mu-P_{1234}\Si)^*
\sum_i x_i - P_{123} \sum_{i \neq j} x_i x_j^{-1} -k P_{123} \label{Tf}
\end{equation} where the dependence on $\bpi$ is encoded in $\Si$.
We define the K-theoretic 4-vertex \begin{equation} \V_\vv (\bpi) \coloneq \hat
a (T_\mathrm{cross}) \, \hat a (T_\mathrm{const}) \, \sum_{k=0}^\infty
(-p)^{|K_\mathrm{reg}|} \oint \hat a (T_\mathrm{formal} + k) \prod_{i=1}^k
\frac{\dd x_i}{x_i} \end{equation} where the subscript $\vv$ keeps track of
local coordinates and the integral is defined using Jeffrey-Kirwan
prescription.  In computing iterated residues, we order $\chi = \chi_1
+ \dots + \chi_k$ as in ref.~\cite{Nekrasov:2018xsb}.  The fugacity
$(-p)^{|K_\mathrm{reg}|}$ comes from \cref{fuga}, and it is equal to
$(-p)^k$, up to a constant factor.  Here we neglect an overall
sign, relevant only for the global picture.  \end{definition}

\begin{theorem} For a given $k$, the admissible poles are in
one-to-one correspondence with the Euler characteristics $\chi$ of
solid partitions $K$ with asymptotics $\bpi$, such that $|\chi|=k$.
\end{theorem} \begin{proof} The relevant measure (neglecting $\mu$ terms)
is \begin{equation} \label{mk} m_k = \prod_{i=1}^k \hat a (-P_{1234} \Si^* x_i)
\prod_{i=1}^k \hat a (x_i) \prod_{k \geq i>j \geq 1} \hat a (-P_{1234}
x_i/x_j). \end{equation} Notice that, except for the $\Si$ term, this is
the same as the finite case \cite[Section 2.3]{Nekrasov:2018xsb}.  Let us assume by
induction that the statement is true for $k-1$.  The tree argument from
finite case still works, so the new pole $u_k$ can only grow by $e_k-e_i$,
with $i<k$, from some $u_i$ belonging to some admissible $\chi_{k-1}$.
(Here, as in ref.~\cite{Nekrasov:2018xsb},
$e_k-e_i$ is a positive root of $\mathfrak{sl}(k)$ corresponding to the last factor in \cref{mk}.) 
Let us work with $Q$ unconstrained first.  There's one new case, which is
when the new box borders $\Si$.  In this case there are two poles (one
from $u_i \in \chi_{k-1}$ and one from some $\si \in \Si$) and one zero
(from $q_a^{-1}\si$ for some $a \in \{1,2,3,4\}$), therefore we still get
a simple pole and the case is admissible.  This means that $\chi_k$ is a
partition in the background of $\Si$, which is what we wanted to prove.
\end{proof}

\begin{theorem} With the ordering for $\chi$ as in \cref{vtx-res},
let's define the sign \begin{equation} s_\vv (\chi)
\coloneq (P_{123}\sum_{i<j} \chi_i \chi^*_j)_0 \end{equation} where the
subscript $0$ denotes unmovable part.
Define the measure \begin{equation} M_\vv (K) \coloneq
(-1)^{s_\vv (\chi)} \, \hat a (T + T_\mathrm{cross} + T_\mathrm{const}).
\end{equation} Then the following holds: \begin{equation} \V_\vv (\bpi)
= \sum_{K \text{ending on}\,\bpi} (-p)^{|K_\mathrm{reg}|} M_\vv (K).
\end{equation} \end{theorem}

\begin{proof} We can prove the relation at each pole, by analyzing
the terms contributing a sign in the residue, as in the finite case
\cite[Section 2.4.2]{Nekrasov:2018xsb}.
The only term in the measure involving $x^{-1}$ is $-P_{123}x_i/x_j$; by
applying induction on the size of $\chi$, we get the thesis.  \end{proof}

\subsection{Examples}

\subsubsection{Vertex with one leg}

Consider the vertex with one non-trivial leg $\pi_1$.  Let $K$ be any
solid partition with asymptotics $\pi_1 = \pi_1 (q_2,q_3,q_4)$ along
direction $q_1$, and trivial asymptotics otherwise.  Let $\chi = K -
\frac{\pi_1}{P_1}$.  Then we have \begin{equation} T = \left(1-\mu
-P_{1234} \frac{\pi_1}{P_1}\right)^* \chi -P_{123}\chi \chi^*.
\end{equation}

This case can be checked against K-theoretic quasimap counts, and this
is done in a companion paper \cite{Piazzalunga:2023qik}.
Here we only write a formula for the simplest example.

\begin{conjecture}
The one-leg vertex with one-box
asymptotics ($\pi_1=1$) takes the form \begin{equation}
\frac{\V_\vv (\square,\varnothing,\varnothing,\varnothing)} {\V_\vv
(\varnothing,\varnothing,\varnothing,\varnothing)} = \sum_{n \geq 0} p^n
\mu^{-n/2} \prod_{k=1}^n \frac {1 - \mu q_1^k} {1 - q_1^k}. \end{equation}
See also ref.~\cite[Lemma~6.5.1]{Monavari:2022umi}
\end{conjecture}

\section{Theory on the edge}

Let $\Sigma$ be a two dimensional Riemann surface, with metric
$g_{\Sigma}$ and local complex coordinates $z, {\bar z}$, in which the
metric $g_{\Sigma}$ is Hermitian, and, therefore, Kähler.  We also fix a
triplet of line bundles $L_{a}$, such that the tensor product of all three
bundles equals the canonical bundle: \begin{equation} L_{1} \otimes L_{2}
\otimes L_{3} \approx K_{\Sigma}. \label{eq:3ls} \end{equation} We endow
each of these bundles with the Hermitian connections ${\varpi}_{a} \dd z +
{\bar\varpi}_{a} \dd {\bar z}$, compatible with \cref{eq:3ls} and the metric
on $\Sigma$.  In a local trivialization, a section $s_a$ of $L_a$ has
the norm \begin{equation} {\rho}_{a}(z, {\zb}) | s_{a} |^2 \end{equation}
so that ${\rho}_{1}{\rho}_{2}{\rho}_{3} = g^{z{\zb}}$, and ${\varpi}_{a}
= \frac 12 {\rho}_{a}^{-1} {\partial} {\rho}_{a}$, ${\bar\varpi}_{a} =
\frac 12 {\rho}_{a}^{-1} {\bar\partial} {\rho}_{a}$.

We are going to define a two-dimensional cohomological field theory,
whose fields are the $\grp{U}(k)$-gauge field $A_{z} \dd z+ A_{\bar z} \dd {\bar z}$,
a triplet of $L_a$-twisted complex adjoint scalars $B_{a}$ and their
$L_{a}^{-1} = {\bar L}_{a}$-twisted conjugates $B_{a}^{\dagger}$, and
a scalar field $I$ valued in the fundamental representation of $\grp{U}(k)$
(i.e., a section of a rank $k$ complex vector bundle associated with
the principal $\grp{U}(k)$ bundle in which $A$ is a connection).

These fields are constrained by the equations, which are the two
dimensional generalization of the equations describing the Hilbert
scheme of points on ${\BC}^{3}$ \cite{Nekrasov:2004vv, Nekrasov:2005bb,
Nekrasov:2009j, Nekrasov:2017cih}: \begin{equation} \begin{aligned}
D_{\bar z} B_{a} + {\ve}_{abc} [ B_{b}^{\dagger}, B_{c}^{\dagger}] &=
0, \qquad a = 1,2,3 \\ D_{\bar z} I &= 0 \\ - g^{z{\zb}} F_{z{\bar z}} +
II^{\dagger} + \sum_{a=1}^{3} {\rho}_{a}^{-1} [ B_{a}, B_{a}^{\dagger}] &=
r \cdot 1. \end{aligned} \label{eq:2dhilb3} \end{equation} Here $D_{\bar z}
B_{a} = {\partial}_{\bar z} B_{a} + {\bar\varpi}_{a} B_{a} + [ A_{\bar z}
, B_{a}]$, and $D_{\bar z}I = {\partial}_{\bar z} I +  A_{\bar z} I$.
The space of solutions to \cref{eq:2dhilb3} is to be modded out by the
group of $\grp{U}(k)$ gauge transformations.  The corresponding moduli space
is a disjoint union of spaces \begin{equation} {\cM}(k) = \coprod_{p
\in {\BZ}} {\cM}_{p}(k) \end{equation} with $p$ being the first Chern
class of the gauge bundle \begin{equation} p = \frac{1}{2\pi \ii}
\int_{\Sigma} \tr F_{A}. \end{equation} For the purposes of this paper,
we shall only need to consider the case of ${\Sigma} = S^2$, with
the metric $g_{\Sigma}$ having a $\grp{U}(1)$ isometry (the round metric
on a sphere is one such example).  Imagine the geometry of a long
cylinder that is capped at the ends by two hemispheres.  In the long
flat region, where $\varpi = 0$, we can look for specific solutions
of \cref{eq:2dhilb3}, namely, the $z,{\bar z}$-independent $B_{1},
B_{2}, B_{3}$ and $B_{4} \equiv A_{\bar z}$.  Then \cref{eq:2dhilb3}
reduce to the familiar equations \begin{equation} \begin{aligned}
[B_{4}, B_{a}] + {\ve}_{abc} [ B_{b}^{\dagger}, B_{c}^{\dagger}] &=
0, \qquad a = 1,2,3 \\ B_{4} I &= 0 \\ II^{\dagger} + \sum_{i=1}^{4}
[ B_{i}, B_{i}^{\dagger}] &= r \cdot 1 \end{aligned} \label{eq:hilb3}
\end{equation} describing the Hilbert scheme of points on ${\BC}^3$.
Towards the caps, the derivative terms become important.

The fields and the equations, together with the gauge symmetry, make up
the field content of twisted ${\cN}=(2,2)$ supersymmetric gauge theory in
two-dimensions.  Unlike the generic theory with four supercharges, which
has only $A$-type or $B$-type twists, this theory admits a variety of
twists due to the extended supersymmetry of its core component.  Namely,
our theory is ${\cN}=(8,8)$ $\grp{U}(k)$ super-Yang--Mills theory, whose symmetry
is broken down to ${\cN}=(2,2)$ by coupling to the fundamental chiral
multiplet, whose complex scalar component is simply $I$.  The ${\cN} =
(8,8)$ part has an $\grp{SO}(8)$ $R$-symmetry group, which partly survives the
coupling to the fundamental chiral, allowing for a variety of twisting
$R$-symmetries.

Our theory can be canonically lifted to a three-dimensional theory
with the same field content, except that one of the real scalars in the
vector multiplet becomes the third component of the gauge field.  It is
this three-dimensional theory that we use in this paper in defining the
edge contributions.

Let us now analyze the solutions to \cref{eq:2dhilb3} in the case
of $\Sigma = S^2$.  First, the norm squared of the first equations
\begin{multline} 0 = \sum_{a=1}^{3} \int_{S^2} \dd^{2}z\, {\rho}_{a}^{-1}
\tr \left( D_{\bar z} B_{a} + {\ve}_{abc} [ B_{b}^{\dagger},
B_{c}^{\dagger}] \right) \left( D_{z} B_{a}^{\dagger} - {\bar\ve}_{abc}
[ B_{b}, B_{c}] \right) = \\ \sum_{a=1}^{3} \int_{S^2} \dd^{2}z\,
{\rho}_{a}^{-1} \, \tr \left( D_{\bar z} B_{a}  D_{z} B_{a}^{\dagger}
\right) + \sum_{b<c}\int_{S^2} \dd^{2}z\, \sqrt{g}\, {\rho}_{b}{\rho}_{c}
\, \tr [ B_{b}, B_{c} ] [B_{b}, B_{c}]^{\dagger} \end{multline}
where we used the identity ${\rho}_{1}{\rho}_{2}{\rho}_{3} \sqrt{g}
= 1$, and \begin{equation} \int_{S^{2}} \bar\partial \tr \left( B_{1}
[ B_{2}, B_{3}] \right) = 0 \end{equation} with the understanding that
$\tr \left( B_{1} [ B_{2}, B_{3}] \right)$ is a $(1,0)$-form on $\Sigma$
(being a section of the canonical bundle).  Thus, \cref{eq:2dhilb3} imply
\begin{equation} D_{\zb} B_{a} = 0, \quad D_{\zb} I = 0, \qquad a =
1, 2, 3 \label{eq:holoeq} \end{equation} and \begin{equation} [ B_{b},
B_{c} ] = 0, \qquad 1 \leq b < c \leq 3. \label{eq:commeq}
\end{equation} The operator ${\nabla}_{\zb} = {\bar\partial}_{\zb} +
A_{\zb}$ defines the structure of a holomorphic rank $k$ bundle $\cE$
over ${\BP}^{1}$, of which $I$ is a holomorphic section, while the
operators $B_{a}$ are the commuting holomorphic twisted Higgs fields:
\begin{equation} I \in H^{0} \left( {\BP}^{1}, {\cE} \right), \quad
B_{a} \in H^{0} \left( {\BP}^{1}, {\cE} \otimes {\cE}^{*} \otimes
L_{a} \right). \end{equation} Now, to proceed algebro-geometrically
we would like to replace the last equation in \cref{eq:2dhilb3}
by an $r$-dependent stability condition, so that instead of solving
the last equation in \cref{eq:2dhilb3} we divide the space of stable
solutions to \cref{eq:holoeq} by the group ${\cG}_{\BC}$ of $\grp{GL}(k,
{\BC})$ gauge transformations \begin{equation} \left( B_{1}, B_{2},
B_{3}, A_{\zb}, I \right) \mapsto \left( g^{-1}B_{1}g, g^{-1}B_{2}g,
g^{-1}B_{3}g, g^{-1}A_{\zb}g+g^{-1}{\bar\partial}_{\zb} g,  g^{-1}I
\right). \end{equation} We can now fix the gauge $A_{\zb} =0$ on the
northern and on the southern hemispheres $H_{\pm}$ with the transition
function $h(z)$ being holomorphic on the intersection $H_{+} \cap H_{-}
= {\BC}^{\times}$.  Grothendieck's theorem allows us to find a conjugacy
class of $h$ in the form of a diagonal matrix \begin{equation} h(z) =
\mathrm{diag} \left( z^{p_{1}}, \dots, z^{p_{k}} \right) \end{equation}
where $p_{i} \in {\BZ}$, $p_{1} \geq p_{2} \geq \dots \geq p_{k}$,
and \begin{equation} p_{1} + p_{2} + \dots + p_{k} = p. \end{equation}
In this gauge the rest of the fields are holomorphic on $H_{\pm}$
respectively, with the identifications \begin{equation} I_{+}(z) = h(z)
I_{-}(z)\, , \quad B_{a,+}(z) = {\ell}_{a}(z) h(z) B_{a, -}(z) h(z)^{-1}\,
, \qquad a = 1, 2, 3 \label{eq:glu2} \end{equation} where ${\ell}_{a}(z) =
z^{-l_{a}}$ are the transition functions of the holomorphic bundles $L_a$.
The isomorphism \cref{eq:3ls} implies \begin{equation} l_{1} + l_{2} +
l_{3} = -2. \end{equation} We see that the non-trivial solutions for $I$
lie in the subspace of ${\BC}^{k}$ spanned by the eigenvectors $e_i$ of
$h$ with $p_i \geq 0$.  The corresponding components ${\upsilon}^{i}(z)$,
\begin{equation} I_{+}(z) = \sum_{i=1}^{k} {\upsilon}^{i}(z) e_{i}
\end{equation} are simply some degree $p_i$ polynomials in $z$, so that
$z^{-p_{i}} {\upsilon}^{i}(z)$ is a degree $p_i$ polynomial in $z^{-1}$.
The classification of possible solutions for $B_a$'s is more involved.

\subsection{ADHM-like model for the one-leg theory}

Let us present the matrix quantum mechanics describing the moduli space of
solutions to the vortex equations \cref{eq:2dhilb3}.  We fix two complex
vector spaces $R,P$, of dimensions $r$ and $p$, respectively, endowed
with Hermitian metrics.  The fields of our model are: \begin{multline}
I \in \Hom (\BC, R), \quad B_{A} \in \End (R), \quad A = 1, 2, 3, 4 \\
\gamma \in \Hom (R,P), \quad \beta_a \in \Hom (P, R), \quad a= 1, 2,
3. \label{eq:adhm2var} \end{multline} They are subject to the equations
\begin{equation} \begin{aligned} [ B_{a}, B_{b} ] + {\ve}_{abc} \left(
[ B_{c}, B_{4}] + {\beta}_{c} {\gamma} \right)^{\dagger} &= 0 \, , \\
B_{a} {\beta}_{b} - B_{b}{\beta}_{a} + {\ve}_{abc} \left( {\gamma} B_{c}
\right)^{\dagger} &= 0 \, , \\ \sum_{a=1}^{3} B_{a}^{\dagger} {\beta}_{a}
+ B_4 \gamma^{\dagger} &= 0 \, , \\ \sum_{A=1}^{4}
[ B_{A}, B_{A}^{\dagger} ] + II^{\dagger} + \sum_{a=1}^{3} {\beta}_{a}
{\beta}_{a}^{\dagger} - {\gamma}^{\dagger}{\gamma} &= {\zeta}_{\BR}
\cdot 1_{R} \end{aligned} \label{eq:adhm2eq} \end{equation} and to the
equivalence relation \begin{equation} \left( I ; B_{A} ; {\beta}_{a}
; {\gamma} \right) \mapsto \left( g^{-1} I ; g^{-1} B_{A} g ; g^{-1}
{\beta}_{a} ; {\gamma} g \right), \quad g \in \grp{U}(r). \label{eq:adhm2sym}
\end{equation}

\subsection{Connection to cigar partition function}

One can interpret the $1$-leg vertex function of the Magnificent Four
theory as the cigar partition function \cite{ND:2023} of the $(2+1)$-dimensional
gauged linear sigma model with the field content described
in the previous subsections. The latter is expected to obey a system
of difference equations, forming part of the rich algebraic structure
hidden in the full $4$-vertex.

\section{Edge}

Recall that the product $P_{1234}$, just like $q_a$ for $a=1,2,3,4$,
depends on the choice of $\vv \in \D_0$.

\begin{definition} Fix a reference $\vv$
and direction $e=1$.  Let
\begin{equation}
\label{edge1/2} \cT_ \ed = \left( 1 - \mu -P_{1234} \sum_{ a \neq e}
\frac{ \lambda_{ae}}{P_{ae}} \right)^* \frac{\pi_{e,\mathrm{reg}}}{P_e} - P_{123}
\frac {\pi_{e,\mathrm{reg}}}{P_e} \frac{\pi^*_{e,\mathrm{reg}}}{P^*_e} - \sum_{\substack{a
\neq e, b \neq e\\a \neq b}} P_{abe}
 \frac{\lambda_{ae}}{P_{ae}} \frac{\lambda^*_{be}}{P^*_{be}}.
\end{equation}
\end{definition}

\begin{lemma}
$\cT_\ed$ is a square root of \cref{edge2}.
$T_\ed=\sum_{\vv \in
\ed} \cT_\ed$ is a movable Laurent polynomial.
\end{lemma} \begin{proof} We take
for simplicity  $e=1$ and focus on the non-constant part of
\cref{edge1/2} \begin{equation} \cT = \left( 1- \mu - P_{1234}
\frac{\Sigma}{P}\right)^* \frac{\chi}{P}-P_{123}\frac{\chi}{P}
\frac{\chi^*}{P^*} \end{equation} where $\chi=\chi_{\pi_e}$,
$\Sigma = \Sigma_e = \pi_{e,1}-\pi_{e,2}+\pi_{e,3}$ and $P=P_e$.
By induction, add a box $\xi$ to $\chi$.  We have $\delta T =
\delta  \cT +\delta  \bar {\cT}$, where bar means `evaluated at the
other vertex', and \begin{equation} \delta \cT = \left(1-P_{1234}
\frac{\Sigma}{P}\right)^* \frac{\xi}{P}-P_{123}\frac{1}{PP^*} -
P_{1234} \frac{\chi}{P}\frac{\xi^*}{P^*}. \end{equation} We work
up to movable terms, and compute \begin{equation} \frac{P_{123}+
\bar P_{123}}{(1-q_1)(1-q_1^*)} =1. \end{equation} For any $m =
q_2^{\alpha-1} q_3 ^{\beta-1}q_4^{\gamma-1}$, we define \begin{equation}
[m] \coloneq \frac1{1-q_1} (\bar m-q_1 m). \end{equation} By introducing the
parallelepiped $\Pi=\Pi(\xi)$ and decomposing, we get \begin{equation}
\delta T = -P_{1234} \frac{\Sigma\setminus \Pi_2}{P} \frac{\xi^*}{P^*}
-P_{1234} \frac{\chi \setminus \Pi_1}{P} \frac{\xi^*}{P^*} -P_{1234}
\frac{\Pi}{P} \frac{\xi^*}{P^*} +\text{c.c.} - 1+[\xi] \end{equation} where
c.c.\ means bar.   We compute $[P_{234} (\Sigma\setminus \Pi_2) \xi^* ]
=0=[P_{234} (\chi\setminus \Pi_1) \xi^* ]$, so that \begin{equation}
\delta T = -1+[\xi] - [P_{234} \Pi \xi^* ] \end{equation} and since
$[P_{234} \Pi \xi^* ] = -[\xi]+ 1$ we have $\delta T=0$.  \end{proof}

\begin{definition} The edge measure is \begin{equation} \E_\ed (\lambda)
\coloneq (-p)^{-f_\ed (\pi_{\ed,\mathrm{reg}}) - |\pi_{\ed,\mathrm{reg}}| f_\ed (\tn)} \ee^{-t_\ed
|\pi_{\ed,\mathrm{reg}}|} \,\hat a (T_\ed) \end{equation} with fugacities
determined by the relevant summands in \cref{charges}. \end{definition}

\begin{remark} The unmovable part of \begin{equation} \frac{q_1
P_{23}-\bar P_{23}}{1-q_1} \end{equation} is $-1$, so that
\begin{equation} \mathrm{const} \coloneq k\frac{q_1 P_{23}-\bar P_{23}}{1-q_1} +
k \end{equation} is a movable Laurent polynomial.  \end{remark}

\section{Theory on a face}

Now let us do a similar exercise in four and five dimensions.  We start
with a complex Kähler surface $S$, with Kähler metric $g_{S}$ and local
holomorphic coordinates $z,w$.  We also need to fix the line bundles
$L_{1}$, $L_{2}$ over $S$, such that \begin{equation} L_{1} \otimes L_{2}
= K_{S}. \end{equation} This choice is similar \cite{Witten:1994cg} to the
choice of the so-called ``basic classes'' of Donaldson--Kronheimer--Mrowka.

The fields of our theory are the gauge field $A$, the two adjoint-valued
complex scalars $B_{1}$, $B_{2}$, twisted by the line bundles $L_{1}$ and
$L_{2}$, respectively, a pair $(I,J)$ of fundamental and anti-fundamental
scalar fields, with $J$ twisted by $K_{S}$, and a pair $({\Upsilon},
{\Psi})$ of fermionic fundamental and anti-fundamental scalar fields,
with $\Psi$ twisted by $K_S$.

Our fields are again constrained by a set of elliptic (modulo gauge
symmetry) equations: \begin{equation} \begin{aligned} D_{\bar z} B_{1} +
D_{w} B_{2}^{\dagger} &= 0, \\ D_{\bar w} B_{1} - D_{z} B_{2}^{\dagger}
&= 0, \\ D_{\bar z} {\Upsilon}  + D_{w} {\Psi}^{\dagger} &= 0, \\
D_{\bar w} {\Upsilon} - D_{z} {\Psi}^{\dagger} &= 0, \\ D_{\bar z}
I  + D_{w} J^{\dagger} &= 0, \\ D_{\bar w} I - D_{z} J^{\dagger} &= 0,
\\ F_{zw} + [B_{1}, B_{2} ] + IJ &= 0, \\ - F_{z{\bar z}} - F_{w{\bar
w}} + II^{\dagger} - J^{\dagger}J + {\Upsilon}{\Upsilon}^{\dagger} -
{\Psi}^{\dagger}{\Psi} + \sum_{a=1}^{2} [ B_{a}, B_{a}^{\dagger}] &=
r \cdot 1. \end{aligned} \label{eq:2dhilb2} \end{equation} The middle
equations in \cref{eq:2dhilb2} can be more invariantly stated as
\begin{equation} {\bar\partial}_{\bar A} I + {\bar\partial}_{\bar
A}^{\dagger} J^{\dagger} = 0 \end{equation} where $J^{\dagger}$ is
naturally viewed as a $(0,2)$-form valued in the same vector bundle,
as $I$.

Our theory is, naturally, a twisted ${\cN}=2$ theory in four dimensions,
which is obtained from a twisted ${\cN}=4$ theory by coupling it to a
(twisted) hypermultiplet in the fundamental representation, and reversed
statistics twisted hypermultiplet in the fundamental representation
(such hypermultiplets naturally occur in theories with negative
branes \cite{Dijkgraaf:2016lym}).

Of course, four-dimensional theory with ${\cN}=2$ supersymmetry with
matter hypermultiplets both in the adjoint and fundamental representations
is strongly coupled in the ultraviolet, and the localization
computations reducing path integral to the semi-classical analysis
are not valid. Fortunately, the reversed statistics hypermultiplet
cancels the contribution of the fundamental hypermultiplet to the beta
function. It is this reversed statistics hypermultiplet which is coupled
to the $\mu$-parameter of the magnificent four theory.

The gauge bundle now can have both $c_1$ and $c_2$ and these result in
the nontrivial edge and vertex contributions in the localization approach.

Our theory canonically lifts to five dimensions, which is the version
used in our paper.

\section{Face}

\begin{definition}
Given a Young diagram $\lambda$ and a box $(i,j)$ in it,
we define its arm and leg lengths as
$\ell = \lambda_j - i$, $\ra = \lambda^t_i - j$,
where $\lambda^t$ denotes the transposed diagram.
\end{definition}

\begin{definition}
Fix reference directions $a=1$ and $b=2$ for $\ff\in\D_2$.
Following ref.~\cite{Nakajima:lectures},
let \begin{equation}
\label{face} \cT_\ff = -\mu^* \frac {\lambda_{12}}{P_{12}} + \sum_{\Box
\in \lambda_{12}} \frac {q_3^{\ell +1} q_4^{-\ra}} {P_{12}^*}
\end{equation} where $\ell$ and $\ra$ are the leg and arm lengths of
$\Box \in \lambda$.
\end{definition}

\begin{lemma} $\cT_\ff$ is a square root of \cref{face2}.
$T_\ff = \sum_ {\vv \in \ff} \cT_\ff$ is a movable Laurent
polynomial.\end{lemma}

\begin{proof} The function \begin{equation} s_{\alpha,\beta} \coloneq \sum_{\vv
\in \ff} \frac {q_3^\alpha q_4^\beta} {(1-q_1^{-1})(1-q_2^{-1})}
\end{equation} is a Laurent polynomial for all $\alpha, \beta \in \BZ$
because it is the equivariant Euler characteristic of the product of
two line bundles over a compact surface. Then \begin{equation} f_2
(\ra,\ell) = \sum_{\vv \in \ff} \frac {q_3^{\ra+\ell+1} (q_1q_2)^{\ra}}
{(1-q_1^{-1})(1-q_2^{-1})} \end{equation} is a Laurent polynomial, being
of the form $s_{\alpha,\beta}$.  It is also movable for all $\ra,\ell
\geq 0$, since $q_3^{(v)}=q_3 q_1^{p(v)} q_2^{q(v)}$ and $h=\ra+\ell+1
>0$. \end{proof}

\begin{definition} The face measure is \begin{equation} \F _\ff \coloneq
(-p)^{g_\ff (\lambda_\ff) - \frac{1}{24} |\lambda_\ff| c_{2,\ff} +
g_\ff (\lambda_\ff,\tn) + \frac12 |\lambda_\ff| g_\ff (\tn)} \ee^{\tilde g_\ff
(\lambda_\ff) + |\lambda_\ff| \tilde g_\ff (\tn) + \frac12 |\lambda_\ff|
A_{002}} \, \hat a (T_\ff) \end{equation} with fugacities determined by
the corresponding summands in \cref{charges}. \end{definition}

\begin{remark} Upon enforcing the CY condition, the index of Dirac
operator gives \begin{equation} \begin{aligned} \# s_{\alpha,\beta}
&= \frac12 \int_S \left[ \left(\alpha-\frac12\right) c_1(\sL_3) +
\left(\beta-\frac12\right) c_1(\sL_4)\right]^2 - \frac\sigma8 \\ &=
\frac18 (S.S-\sigma)+\frac12 \beta(\beta-1) S.S + \frac12 (\beta-\alpha)^2
\sL_3.\sL_3 +S.\sL_3 (\beta-\frac12)(\beta-\alpha) \end{aligned}
\end{equation} whose integrality implies \begin{equation} \label{integ}
S.S -\sigma=0 \mod 8, \quad S.\sL_3 + \sL_3.\sL_3 =0 \mod 2. \end{equation}
Here $S.S = \int_S c_1(S)^2$, $S.\sL_3 = \int_S c_1(S) c_1(\sL_3)$, and
$\sL_3.\sL_3 = \int_S c_1(\sL_3)^2$.  The first equation in \cref{integ}
follows from Hirzebruch signature theorem \begin{equation} 2\chi+3
\sigma = S.S \end{equation} while the second is equivalent to $A_{110}$
and $\int_{S} e({\cN})$ being even (see \cref{A110-even},
where $\cN$ is defined).\end{remark}

\subsection{Virtual dimension}

Recently, there has been some progress \cite{Bae:2022pif, Bae:2024bpx}
on surface counting in the CY4 setting.
In order to facilitate comparison and future work,
let us compute the number of monomials in the (movable)
Laurent polynomial $T_\ff$.

\begin{lemma} With definitions as in \cref{charges},
if we decompose the Laurent polynomial $T_\ff$ as $P_1 + \tm P_2$,
then the identities
\begin{equation}
\# P_1 \eqcolon \operatorname{vdim} = g(\lambda) - \frac12 |\lambda|^2 A_{110}
- \frac18 |\lambda| (\sigma+\tilde \sigma)
\end{equation}
as well as
\begin{equation}
\# P_2 =
-\frac12 |\lambda| g(\tn) - g(\lambda,\tn) - g(\lambda)
+\frac18 |\lambda| (\sigma + \tilde \sigma)
\end{equation}
hold true for any compact face $\ff$ in a toric CY4.
\end{lemma}

\begin{proof}
Up to conjugation, we can write \cref{face} as
\begin{equation}
\cT_\ff = \sum_{\Box \in \lambda} \frac
{q_3^{\ell +1} q_4^{-\ra} - \tilde\mu q^{-\tn} q_3^{1-i} q_4^{1-j}}
{P_{12}^*}.
\end{equation}
We apply twice the remark above:
\begin{equation}
\# P_1 = \frac12 \sum_{\Box \in \lambda}
\left[ \left( \ell+\frac12 \right) \sL_3
- \left( \ra+\frac12 \right) \sL_4 \right]^2
- \frac{\sigma|\lambda|}8.
\end{equation}
The integrand instead depends on $\mu$, and therefore on the
the restriction to $S$ of the line bundle $L = \sum_{\cc\in\Delta_3}
\tn_\cc L_\cc$, with $L_\cc$ the line bundle associated to cell $\cc$:
\begin{equation}
\# P_2 = - \frac12 \sum_{\Box \in \lambda}
\left[ \left( i-\frac12\right) \sL_3 + \left(j-\frac12\right) \sL_4 + L \right]^2
+ \frac{\sigma|\lambda|}8.
\end{equation}

For the purpose of this proof, we define
$\tilde \sigma_\ff \coloneq \frac13 (A_{200} + A_{020})$, and
\begin{equation}
\hat g (\lambda) \coloneq \sum_{\Box \in \lambda} \left[
\left( \frac{\ell(\ell+1)}{2} + \frac16 \right) A_{200}
+ \left( \frac{\ra(\ra+1)}{2} + \frac16 \right) A_{020}
- \left( \ell\ra + \frac12 (\ra+\ell) + \frac14 \right) A_{110}
\right].
\end{equation}
We claim that $\hat g (\lambda) - g_\ff (\lambda) = -
\frac12 |\lambda|^2 A_{110}$.

We immediately recognize various terms
\begin{equation}
- \frac12 |\lambda| L^2 = - \frac12 |\lambda| g_\ff (\tn)
\end{equation}
as well as
\begin{equation}
- \sum_{\Box \in \lambda} [(i-\frac12) \sL_3 + (j-\frac12) \sL_4)] \cdot L
= - g_\ff (\lambda,\tn)
\end{equation}
and
\begin{equation}
- \frac12 \sum_{\Box \in \lambda} [(i-\frac12) \sL_3 + (j-\frac12) \sL_4)]^2
= - g_\ff (\lambda) + \frac18 \tilde \sigma_\ff |\lambda|.
\end{equation}

From this, we can read off the virtual dimension of a face:
\begin{equation}
\operatorname{vdim} = g(\lambda) - \frac12 |\lambda|^2 A_{110}
- \frac18 |\lambda| (\sigma+\tilde \sigma)
\end{equation}
and of the integrand
\begin{equation}
\# P_2 =
-\frac12 |\lambda| g(\tn) - g(\lambda,\tn) - g(\lambda)
+\frac18 |\lambda| (\sigma + \tilde \sigma).
\end{equation} 
Finally, we get the equality
\begin{equation}
\# T_\ff =
- \frac12 |\lambda|^2 A_{110} - \frac12 |\lambda| g(\tn) - g(\lambda,\tn).
\end{equation}

The last step is to show that $\hat g (\lambda) - g_\ff (\lambda) = -
\frac12 |\lambda|^2 A_{110}$.  We do this in three steps.
First, the identity
\begin{equation}
\sum_{\Box \in \lambda} \ell(\ell+1) - i(i-1) = 0
\end{equation}
follows from the fact that, for every $j$, we have
\begin{equation}
\sum_{\ell=0}^{\lambda_j-1} \ell(\ell+1) = \sum_{i=1}^{\lambda_j} i(i-1).
\end{equation}
Taking the transpose of the diagram, we get a similar identity for
the arm-length.
Second, the identity
\begin{equation}
\frac12 \sum_{\Box \in \lambda} \ra+\ell - (i+j) = - |\lambda|
\end{equation}
follows from computing the terms:
$\sum_{\Box \in \lambda} \lambda_j = \sum_{j=1}^h \lambda_j^2$,
where $h$ is the height of first column;
$\sum_{\Box \in \lambda} j = \sum_{i=1}^{h^t} \frac12 \lambda_i^t (\lambda_i^t+1)$,
where $t$ denotes transpose and $h^t$ is the length of first row of $\lambda$;
$\sum_{\Box \in \lambda} \lambda_i^t = \sum_{i=1}^{h^t} (\lambda_i^t)^2$;
$\sum_{\Box \in \lambda} i = \sum_{j=1}^{h} \frac12 \lambda_j (\lambda_j+1)$.
Finally, we claim the identity
\begin{equation}
\sum_{\Box \in \lambda} \ra\ell+ ij = \frac12 |\lambda| (|\lambda|+1)
.\end{equation}
These three steps together imply $\hat g - g = - A_{110} \frac12
|\lambda|^2$.

Let us prove the claim by induction on the size of $\lambda$.  It is
true for $\lambda=1$.  Assume it is true for $|\lambda|=k$.  Let $\mu$
be the partition obtained from adding a box $(p,q)$ to $\lambda$.  Let us
consider two types of boxes $(i,j) \in \lambda$: first, the boxes with
$j=q$ have $\ell_\mu = \ell_\lambda + 1$; second, the boxes with $i=p$
have $\ra_\mu = \ra_\lambda + 1$.  Correspondingly, we can write
\begin{equation}
\sum_{\Box \in \mu} \ra_\mu \ell_\mu + ij =
\sum_{\Box \in \lambda} \ra_\lambda \ell_\lambda + ij
+ \sum_{j=q,i=1,\ldots,\lambda_q} \ra_\lambda 
+ \sum_{i=p,j=1,\ldots,\lambda_p^t} \ell_\lambda 
+ (\lambda_q+1)(\lambda_p^t + 1)
\end{equation}
where the last term comes from the box $(p,q)$ itself.  By the induction
hypothesis, the first term equals $\frac12 k (k+1)$, while the sum of
the last three terms equals $k+1$.  A way to see this is to decompose
$\lambda$ as the sum of three terms: the rectangle generated
by $(p,q)$, which has $\lambda_q \lambda^t_p$ boxes; anything above
it, which has $\lambda_1^t + \dots + \lambda^t_{\lambda_q} - (q-1)
\lambda_q$ boxes; and anything to its right, which has $\lambda_1 +
\dots + \lambda_{\lambda^t_p} - (p-1) \lambda_p^t$ boxes.
\end{proof}

\section{Perturbative matters}

\begin{definition} Define \begin{equation} T_6 \coloneq \tm/\tnu \sum_{\vv\in
\D_0} \frac{1-q^{-\tn}}{P_{1234}}. \end{equation} \end{definition}

\begin{lemma} Fix any reference vertex $\vv \in \D_0$.  Then $T_6$ is
a Laurent polynomial in $\tm$ and the local variables $q_a$'s at $\vv$.
It is movable since it is multiplied by $\tm$.  By the condition $Q=1$,
it is a square root of $T^2_6$ in \cref{ts6}.  \end{lemma} \begin{proof}
Set $\tnu=1$ without loss of generality.  Up to complex conjugation,
let us write \begin{equation} T_6 = - \tm^* \sum_{\vv \in \D_0} \left(
\tK - q^{\tn} K \right) \end{equation} and observe that in the RHS
of \begin{multline} \label{t6} \sum_{\vv \in \D_0} (\tK - q^{\tn} K) =
\sum_{\vv \in \D_0} (\tK_\mathrm{reg} - q^{\tn} K_\mathrm{reg}) + \sum_{\ed \in \D_1}
\sum_{\vv \in \ed} \frac{\tilde \pi_{e,\mathrm{reg}} - q^{\tn} \pi_{e,\mathrm{reg}}}{P_e}
\\ +\sum_{\ff \in \D_2} \sum_{\vv \in \ff} \frac{\tilde \lambda_{ab,\mathrm{reg}}
- q^{\tn} \lambda_{ab}}{P_{ab}} +\sum_{\cc \in \D_3} \sum_{\vv \in \cc}
\frac{\tn_{abc}}{P_{abc}} \end{multline} the first sum is clearly finite,
in the second and third sums each summand in the differences is finite
once summed over vertices in the appropriate edge or face, while the
last term is a sum of integrals over compact cells in the $\beta \to
0$ limit.  \end{proof}

\begin{remark} One can reabsorb all the terms in \cref{t6} except the
last one by redefining the vertex, edge and face, replacing schematically
the term $\mu^* X_\mathrm{reg}/P$ by $\tilde \mu^* \tilde X_\mathrm{reg}/P$ for $X$
a solid, plane or ordinary partition respectively and $P$ the appropriate
denominator.  \end{remark}

\begin{definition} Let \begin{equation} \CC \coloneq (-p)^{-h_4 (\tn) + \frac12
\frac{1}{24} c_{2,2,0} (\tn) + \frac{\zeta(3)}{(2\pi)^3} c_{3,1,0}}
\ee^{-h_3 (\tn) + \frac{1}{24} c_{2,1,1} (\tn) - h_2(\tn) - h_1 (\tn)}
\,\hat a (T_6) \end{equation} where the fugacity comes from the relevant
summands in \cref{charges}.  \end{definition}

Define through $\zeta$-function regularization \begin{equation} \hat
a \left( \frac{1-\tm}{P_{1234}} \right) \coloneq \tm^{-\frac12 \zeta_4} PE
\left( \frac{1-\tm^*}{P_{1234}} \right) \end{equation} where $\zeta_4 =
\sum_{n_1,n_2,n_3,n_4=0}^\infty 1$, and let \begin{equation} Z_\mathrm{pert}
\coloneq \prod_{\vv \in \D_0} \hat a \left( \frac{1-\tm}{P_{1234}} \right).
\end{equation} This is the only infinite product left after reshuffling
terms.  We don't need to worry about its sign, as it multiplies the
whole partition function.  The rest of the instanton configuration
is given by the movable Laurent polynomial \begin{equation} I = T_6 +
\sum_{\vv \in \D_0} T_\vv + \sum_{\ed \in \D_1} T_\ed + \sum_{\ff \in
\D_2} T_\ff. \end{equation}

\section{Conclusions and a conjecture}

We presented a construction, motivated by gauge theory, which allows us
to define the K-theoretic vertex of fourfolds for arbitrary asymptotics,
in a combinatorial fashion and with concrete signs.

\begin{conjecture}
There exist sign choices $s_\ff (\lambda^{(\ff)})$ and $s_\ed (\pi^{(\ed)})$
such that \cref{dt} reads \begin{equation} \label{conj}
\Z = \sum_{P\in\mathsf P} \CC \prod_{\vv \in \D_0} \V_\vv \prod_{\ed \in
\D_1} (-1)^{s_\ed (\pi^{(\ed)})} \E_\ed \prod_{\ff \in \D_2}
(-1)^{s_\ff (\lambda^{(\ff)})} \F_\ff \end{equation} where the sum is
over all collections of partitions \begin{multline} \mathsf P = \{ P =
(\tn^{(1)},\dots,\tn^{(|\D_3|)}; \lambda^{(1)},\dots,\lambda^{(|\D_2|)};
\pi_\mathrm{reg}^{(1)},\dots,\pi_\mathrm{reg}^{(|\D_1|)}) \mid \\ P \text{ is
the profile of a fixed point } \{ \tK^\vv, \, \vv \in \D_0 \} \}.
\end{multline} 
\end{conjecture}

While the signs for the vertex $\V$ are determined in the
present work, and the corresponding part of the partition function is
completely fixed, the interaction terms between edges and faces, which
are overall from the viewpoint of a vertex, as well as the signs for edge
and face are not.
We plan to
address this shortcoming, as well as present some examples, in part II.

Once we fix $X$, the partition function $\Z$ depends on the fugacities
(see \cref{fuga}), the mass $\tilde \mu$, and four $\Omega$ background
parameters $q_1$, $q_2$, $q_3$, $q_4$ with product $Q=1$ (one can choose
a reference vertex and express the local variables at the other vertices
in terms of it).  It may be possible to prove \cref{conj} using derived
algebraic geometry \cite{Borisov:2015vha}.

\subsection*{Acknowledgements}
NP thanks M.de Cataldo, L.Cassia, E.Diaconescu, S.Franco, M.Kool,
V.Saxena for discussions.
The research of NP was supported by the US Department of Energy
under grant DE-SC0010008.
We also thank the anonymous referees for their detailed feedback.

\printbibliography
\end{document}